%% file: nf2_baryon.tex
\documentclass[aps,prd,showpacs, nofootinbib, superscriptaddress, floatfix,preprintnumbers]{revtex4}%
\usepackage{amssymb,hyperref, amsmath}
\usepackage[dvips]{color}
\usepackage[dvips]{graphicx}
\usepackage{bm,graphicx}
\newcommand{\pvec}{\bm{p}}
\newcommand{\xvec}{\bm{x}}

\usepackage{epsf,latexsym,amssymb}
\usepackage{epsfig}
\usepackage{amsmath}            
\preprint{JLAB-THY-08-929,RBRC-764}

\begin{document}
\newcommand{\beq}{\begin{equation}}
\newcommand{\eeq}{\end{equation}}
\newcommand{\tr}{\mbox{tr}\,}

\bibliographystyle{apsrev}
\title{Excited State Nucleon Spectrum with Two Flavors of Dynamical Fermions}

\author{John M. Bulava}
\email{jbulava@andrew.cmu.edu} \affiliation{Department of Physics, Carnegie Mellon University, Pittsburgh, PA 15213}

\author{Robert G. Edwards}
\email{edwards@jlab.org} \affiliation{Thomas Jefferson National
Accelerator Facility, Newport News, VA 23606}

\author{Eric Engelson}
\email{engelson@umd.edu} \affiliation{Department of Physics, University of Maryland, College Park, MD 20742, USA}

\author{Justin Foley}
\email{jfoley@andrew.cmu.edu} \affiliation{Department of Physics, Carnegie Mellon University, Pittsburgh, PA 15213}

\author{B\'alint Jo\'o}
\email{bjoo@jlab.org}
\affiliation{Thomas Jefferson National Accelerator Facility,
Newport News, VA 23606}

\author{Adam Lichtl}
\email{alichtl@bnl.gov} \affiliation{RIKEN-BNL Research Center, Brookhaven National Laboratory, Upton, NY 11973}

\author{Huey-Wen Lin}
\email{hwlin@jlab.org} \affiliation{Thomas Jefferson National
Accelerator Facility, Newport News, VA 23606}

\author{Nilmani Mathur}
\email{nilmani@theory.tifr.res.in} \affiliation{Department of Theoretical Physics, Tata Institute of Fundamental Research, India}

\author{Colin Morningstar}
\email{colin_morningstar@cmu.edu} \affiliation{Department of Physics, Carnegie Mellon University, Pittsburgh, PA 15213}

\author{David G. Richards}
\email{dgr@jlab.org} \affiliation{Thomas Jefferson National
Accelerator Facility, Newport News, VA 23606}

\author{Stephen J. Wallace}
\email{stevewal@umd.edu} \affiliation{Department of Physics, University of Maryland, College Park, MD 20742, USA}

\date{\today}

\begin{abstract}
Highly excited states for isospin $\frac{1}{2}$ baryons are calculated for the first time using lattice QCD with two flavors of dynamical quarks.
Anisotropic lattices are used with two pion masses, $m_{\pi}$ = 416(36) MeV and 578(29) MeV.
The lowest four energies are reported in each of the six irreducible representations of
the octahedral group at each pion mass.  
The lattices used have dimensions 24$^3\times$64,
spatial lattice spacing $a_s \approx$ 0.11 fm and temporal lattice spacing $a_t = \frac{1}{3} a_s$.
Clear evidence is found for a $\frac{5}{2}^-$ state in the
pattern of negative-parity excited states.  This agrees with
the pattern of physical states and spin $\frac{5}{2}$ has been 
realized for the first time on the lattice.

\end{abstract}

\maketitle

\input{intro2.tex}

\input{setup.tex}

\input{algorithm.tex}

\input{conventional2.tex}
\input{scale.tex}
\input{operators2.tex}

\input{results4.mod.tex}

\input{multihadrons.tex}
\input{conclusions.tex}


\bibliography{nf2_baryon}
\end{document}

%% file: intro2.tex
\section{Introduction}

A major goal for lattice quantum chromodynamics (QCD) is the determination of the spectrum of
the excited states of QCD.  This goal is a complement to 
experimental work that studies the hadrons and their excitations and decays.
In recent years, large amounts of data have been collected at Jefferson Laboratory
regarding the spectrum of excitations of nucleons.  The Excited Baryon Analysis Center
aims to analyze the data using the best hadronic models available.\cite{Lee:2006xu,Matsuyama:2006rp}  Lattice
QCD calculations are needed as a means to link this program to the Lagragian
of QCD.

When QCD was formulated as the basic theory
that would explain hadrons and their excited states, it
could not be solved for the mass spectrum from first
principles because of its nonperturbative
nature.  Much effort over the past thirty years has been devoted to 
developing the methods and tools to solve QCD on a lattice.  
Accurate resolution of the excited states of hadrons using lattice QCD has proven difficult.
In Euclidean space, excited state correlation functions decay faster than
the ground state.  At large times, the signals for excited states are swamped
by the signals for lower energy states.  Improved resolution in the temporal 
direction is essential for progress.  An anisotropic lattice where the temporal
lattice spacing is finer than spatial spacings can provide better
resolution while avoiding the increase in computational cost associated with
a similar reduction of all spacings.  The improved resolution must be combined with
two other ingredients.  A large number of operators is required that overlap well with excited states.
The use of variational methods is essential to separate the excited states.

Large sets of baryon operators were developed and projected to the irreducible representations of the octahedral group in Refs. \cite{Basak:2005aq, Basak:2005ir}  Link smearing
and quark smearing were found to both be needed in order to optimize the quality
of the signals that are obtained with the
operators in Ref.~\cite{Basak:2005gi}.
Variational methods were used to determine the spectra of $I = \frac{1}{2}$ and $I=\frac{3}{2}$ 
excited baryons using the quenched approximation in Ref.~\cite{Basak:2007kj}.

In this work, we take another step toward the goal of determining the spectrum
of nucleon excited states by
studying the spectrum of isospin $\frac{1}{2}$ excited baryons in
two-flavor QCD, using $u$ and $d$ quarks that have the same mass.
Results are obtained on 24$^3\times$64 lattices with two values of the
pion mass: 416(36) MeV and 578(29) MeV.  

In the physical spectrum for isospin $\frac{1}{2}$ the lowest three states are the
nucleon, $N$, the Roper resonance, $N^\prime$ ($P_{11}$) and
the opposite-parity $N^*$ ($S_{11})$).  Quenched lattice QCD
calculations~\cite{Sasaki:2001nf,Guadagnoli:2004wm,Leinweber:2004it,Sasaki:2005ap,Sasaki:2005ug,Burch:2006cc}
generally have found a spectrum inverted with respect to experiment, with the $N^\prime$
heavier than the negative-parity $N^*$.  An exception is the calculation of the Kentucky
group~\cite{Mathur:2003zf} that obtained the correct mass ordering
with a pion mass below 400~MeV (after subtracting the effects
of the quenched ``ghosts'').  This has helped to motivate
full QCD simulations where the spectrum can be determined without
unphysical contributions from ``ghost'' states.  Moreover, many
additional excited states have been observed experimentally that should
be reproduced by lattice QCD and full-QCD simulations are needed as a
complement to the experimental searches for new excited states.

Anisotropic techniques have been adopted in lattice
calculations for relativistic heavy quark actions
for the spectrum of charmonium~\cite{Chen:2000ej,Okamoto:2001jb}, for
calculations of the spectrum of
glueballs~\cite{Morningstar:1999rf} and to extract excited baryon states
~\cite{Basak:2005aq,Basak:2005ir,Lichtl:2006dt,Juge:2006gr,Basak:2006ww,Basak:2007kj,Dudek:2007wv}.
Previous results using anisotropic lattices include two-flavor
anisotropic dynamical simulations performed by CP-PACS~\cite{Umeda:2003pj}
and the TrinLat collaboration~\cite{Morrin:2006tf}.  
%

In this work, we report the nucleon spectrum using the interpolating basis from
Refs.~\cite{Basak:2005aq,Basak:2005ir} on two-flavor, anisotropic, Wilson fermion and Wilson gauge configurations.
The action parameters and bare gauge and fermion anisotropies are
tuned such that the gauge anisotropy (as determined from Wilson loop ratios)
and the fermion anisotropy (as determined from the meson dispersion
relation) are both consistent with the desired renormalized anisotropy $a_s/a_t=3$.
Our configurations were generated using the
Chroma~\cite{Edwards:2004sx} HMC code with 
multi-timescale integration.
%

The organization of this paper is as follows: In Sec.~\ref{sec:Setup}, we
discuss the details of the actions used and their parameters.
Then in Sec.~\ref{sec:Algorithm} we discuss the Hybrid Monte
Carlo (HMC) used in this work and show how it is applied to anisotropic
lattices with mass preconditioning.
Section~\ref{sec:conventional} presents results for the conventional
determination of hadron masses and the anisotropy from two-point correlation functions and
Sec.~\ref{sec:scale} presents our procedure and results for
setting the lattice scale in physical units.
Section~\ref{sec:baryon-ops} discusses the construction of large numbers of baryon operators in
the relevant irreducible representations of the octahedral group and demonstrates the
noise suppression that is obtained by smearing both the quark and gauge fields.
Section~\ref{sec:results} presents
results for the $I=\frac{1}{2}$ baryon spectrum
for pion masses of 416 MeV and 578 MeV using $N_f=2$ lattices.
Clear evidence for a spin-$\frac{5}{2}^-$ state is presented.
Some conclusions are presented in Sec.~\ref{Sec:Conclusion}.

%% file: setup.tex
\section{Lattice Actions}\label{sec:Setup}


In this section, we describe the gauge and fermion actions used in
this calculation. For the gauge sector, we use a Wilson anisotropic gauge
action
\begin{eqnarray}\label{eq:aniso_syzG}
S_G^{\xi}[U] &=& \frac{\beta}{N_c\xi_0} \left\{
\sum_{x,s\neq s^\prime}  
\Omega_{{\cal P}_{ss^\prime}}(x)
+ \sum_{x,s}\xi_0^2  
\Omega_{{\cal P}_{st}}(x) \vphantom{\frac{1}{\xi}} \right\},
\end{eqnarray}
where $\Omega_W={\rm Re}{\rm Tr}(1-{\cal P})$ and ${\cal P}$ is the plaquette
\begin{equation}
{\cal P}_{\mu\nu}(x) = U_\mu(x) U_\nu(x+\mu) U_\mu^\dagger(x+\nu) U_\nu^\dagger(x).
\end{equation}
The coupling $g^2$ appears in $\beta=2 N_c/g^2$.
The parameter $\xi_0$ is the bare gauge anisotropy. 
In the fermion sector, we adopt the anisotropic Wilson fermion
action~\cite{Klassen:1998fh}
\begin{eqnarray}\label{eq:anisoW}
S_F^{\xi}[U, \overline{\psi},\psi ]
&=& a_s^3 a_t \sum_{x} \overline{\psi}(x) M_W \psi(x),\nonumber\\
M_W &=&m_0 + \nu_t {\mathcal W}_t +{\nu_s} {\mathcal W}_s , \nonumber \\
\end{eqnarray}
where
\begin{eqnarray}
\mathcal{W}_\mu &=& \nabla_\mu -\frac{a_\mu}{2} \gamma_\mu \Delta_\mu, \nonumber\\
\nabla_\mu f(x) &=& \frac{1}{2a_\mu} \bigg[ U_\mu(x) f(x+\mu) -
    U^\dagger_\mu(x-\mu) f(x-\mu) \bigg], \nonumber\\
\Delta_\mu f(x) &=& \frac{1}{a_\mu^2} \bigg[ U_\mu(x) f(x+\mu) +
    U^\dagger_\mu(x-\mu) f(x-\mu)
- 2f(x) \bigg].
\end{eqnarray}
In terms of dimensionless variables $\hat \psi = a_s^{3/2}
\psi$, $\hat m_0 = m_0 a_t$, ${\hat \nabla}_\mu = a_\mu \nabla_\mu$,
${\hat \Delta}_\mu = a_\mu^2 \Delta_\mu$,
and the dimensionless ``Wilson operator'' $\hat{\mathcal W}_\mu
\equiv \hat \nabla_\mu - \frac{1}{2} \gamma_\mu \hat
\Delta_\mu$, we find that the fermion matrix $M_W$ becomes
\begin{eqnarray}\label{eq:pre-fermion-action}
M_W & = & \frac{1}{a_t} \left\{
 a_t \hat{m_0} + \nu_t \hat{\mathcal W}_t +\frac{\nu_s}{\xi_0}  \sum_s  \hat{\mathcal W}_s
      \right\}.
\end{eqnarray}
Because it is possible to redefine the fields as in Refs.~\cite{Symanzik:1983dc,Symanzik:1983gh}, one
coefficient (either $\nu_t$ or $\nu_s$) is redundant; here we set  $\nu_t = 1$
and $\nu_s=\nu$ for tuning.
For convenience of parameterization, we use the bare gauge and fermion anisotropies, $\gamma_{g,f}$, defined as
\begin{eqnarray}
\gamma_g = \xi_0, \quad
\gamma_f = \frac{\xi_0}{\nu}.
\end{eqnarray}

The parameters $\gamma_g$, $\gamma_f$ and the quark mass $m_0$ require tuning in order to realize the desired renormalization
constraints. The bare gauge and fermion anisotropy parameters $\gamma_g$ and $\gamma_f$ are tuned to obtain the desired renormalized gauge and fermion anisotropies ($\xi_g$ and $\xi_f$): both equal to $a_s/a_t=3.0$.
The renormalized gauge anisotropy ($\xi_g$) can be  determined from the static-quark potential using
Klassen's ``Wilson-loop ratio''~\cite{Klassen:1998ua}:
\begin{eqnarray}\label{eq:WL_ratio}
R_{ss}(x,y) &=&  \frac{W_{ss}(x,y)}{ W_{ss}(x+1,y)}
\xrightarrow{\rm asym} e^{-a_s V_s (y a_s)} ,
  \nonumber \\
R_{st}(x,t) &=&  \frac{W_{st}(x,t)}{W_{st}(x+1,t)}
\xrightarrow{\rm asym} e^{-a_s V_s (t a_t)}, \nonumber\\
\end{eqnarray}
where  $W_{st}$ are the Wilson loops involving the temporal direction, and
$W_{ss}$ are those involving only the spatial directions.
We determine the renormalized gauge anisotropy $\xi_g$ by minimizing~\cite{Umeda:2003pj}
\begin{equation}\label{eq:xiR}
L(\xi_g) = \sum_{x,y}\frac{(R_{ss}(x,y)-R_{st}(x,\xi_g y))^2}
{(\Delta R_s)^2+(\Delta R_t)^2},
\end{equation}
where $\Delta R_s$ and $\Delta R_t$ are the statistical errors of $R_{ss}$ and $R_{st}$. 
A fixed background gauge field in the spatial ``z'' direction is used following the
Schr\"odinger-functional
scheme~\cite{Luscher:1996ug} which allows for a determination of the critical
mass using the PCAC Ward identity.
For more details, see Sec.~IV~B of Ref.~\cite{Edwards:2008ja}.
We determine the renormalized fermion anisotropy $\xi_f$ through the conventional
relativistic meson dispersion relation as will be discussed in 
Sec.~\ref{subsec:SimpleMeson}. 

We find that when $\xi_g=\xi_0=2.38$, $\nu=1$ (or $\xi_f=\xi_0=2.38$),
$\xi_g$ and $\xi_f$ (see Sec.~\ref{subsec:SimpleMeson}) are consistent with 3, given our other choices of parameters. 
The critical mass at these bare parameters is $m_c=-0.41473$.
The $m_0$ parameter within our range of interest has negligible effect on the anisotropies. (Similar results are observed in the three-flavor anisotropic clover action study in Ref.~\cite{Edwards:2008ja}.) We set $m_0$ to $-0.4086$ and $-0.4125$ for our pion-mass study.

%% file: algorithm.tex
\section{Algorithm}\label{sec:Algorithm}

The anisotropic Wilson configurations were generated with the 
Hybrid Monte Carlo (HMC) algorithm~\cite{Duane:1987de}. To increase
the efficiency of the method we employed several techniques such
as Hasenbusch style mass preconditioning~\cite{Hasenbusch:2002ai}, 
the use of multiple timescale integration schemes~\cite{Sexton:1992nu},
chronological inversion methods~\cite{Brower:1995vx} during the 
molecular dynamics, and evolving the temporal links with different
time-steps than the spatial ones~\cite{Morrin:2006tf}. We discuss some of the pertinent
details below:

\subsection{Hybrid Monte Carlo}
The basic technique for gauge generation is a Markov Chain Monte Carlo
method where one moves from an initial gauge configuration to a
successive one by generating a new trial configuration and then
performing an acceptance/rejection test upon the new one. If the trial
configuration is accepted, it becomes the next configuration in
the chain, otherwise the original configuration becomes the next state in the chain.

In order to use a global Metropolis accept/reject step with a
reasonable acceptance rate, the space of states is extended to include
momenta $\pi_{\mu}(x)$ canonical to the gauge links $U_{\mu}(x)$ so
that one may define a Hamiltonian
\begin{equation}
H = \frac{1}{2} \sum_{x,\mu} \pi_{\mu}(x)^\dagger \pi_{\mu}(x) + S(U),
\end{equation}
where $S$ is the action. It is then possible to propose new
configurations from previous ones by performing Hamiltonian Molecular
Dynamics (MD) to get from the initial to the proposed state. Using a
reversible and area preserving MD evolution maintains detailed balance,
which is sufficient for the algorithm to converge. In order to ensure
ergodicity in the entire phase space, the momenta need to change
periodically. This can be accomplished by refreshing the momenta from
a Gaussian heat bath prior to the MD update step.

In order to deal with the fermion determinant, it is standard to use
the method of pseudofermions. One integrates out the Grassman-valued
fermion fields in the action and rewrites the resulting determinant as
an integral over bosonic fields,
\begin{equation}\label{eq:FermPartition}
Z = \int [d\bar{\eta}] [d\eta] e^{-\bar{\eta} \mathcal{D} \eta} =
\det\left( \mathcal{D} \right) = \int [d\phi^\dagger] [d \phi]
e^{-\phi^\dagger \mathcal{D}^{-1} \phi},
\end{equation}
where $\eta$ and $\bar{\eta}$ are the Grassman valued fields,
$\mathcal{D}$ is some Hermitian, positive-definite kernel and $\phi^\dagger$ and $\phi$
are the bosonic pseudofermion fields. Our phase space is thus enlarged
to include also the pseudofermion fields. Like the
momenta, these fields need to be refreshed before each MD step to carry out the
pseudofermion integral.

In the case of a two-flavor simulation, $\mathcal{D}$ is typically of
the form,
\begin{equation}
\mathcal{D} = Q^\dagger Q.
\end{equation}
For the rest of this work $Q$ is an  even-odd preconditioned fermion matrix for an individual flavor of
fermion. In this case $\mathcal{D}$ is manifestly Hermitian and
positive definite, and the integral in Eq.~\ref{eq:FermPartition} is
guaranteed to exist. Furthermore, the pseudofermion fields can easily
be refreshed by producing a vector $\chi$ filled with Gaussian noise
with a variance of $\frac{1}{2}$ and then forming $\phi = Q^\dagger
\chi$.

\subsection{Multiple Time Scale Anisotropic Molecular Dynamics Update}
While any reversible and area-preserving MD update scheme can be used
in the MD step, the acceptance rate is controlled by the truncation
error in the scheme. This manifests itself as a change in the
Hamiltonian, $\delta H$, over an MD trajectory, since we use the
Metropolis acceptance probability
\begin{equation}
P_{\rm acc} = {\rm min}\left( 1, e^{-\delta H} \right) \ .
\end{equation}
We may easily construct a reversible scheme by combining
symplectic update steps $\mathcal{U}_{p}(\delta \tau)$ and
$\mathcal{U}_{q}(\delta \tau)$ which update momenta and coordinates by
a time step of length $\delta \tau$ respectively
\begin{eqnarray}
  \mathcal{U}_{p}(\delta \tau_{\mu}) : \ \left( \pi_{\mu}(x),
  U_{\mu}(x) \right) & \rightarrow & \left( \pi_{\mu}(x) +
  F_{\mu}(x)\delta \tau_{\mu}, U_{\mu}(x) \right), \\
  \mathcal{U}_{q}(\delta \tau_{\mu}) : \ \left( \pi_{\mu}(x),
  U_{\mu}(x) \right) & \rightarrow & \left( \pi_{\mu}(x) , e^{i \pi_{\mu}
  \delta \tau_{\mu}} U_{\mu}(x) \right),
\end{eqnarray}
where $F_{\mu}(x)$ is the MD force coming from the variation of the action with respect to the gauge fields. We emphasize that one may update all the links pointing in direction $\mu$ with a separate step size $\delta \tau_{\mu}$. While this may not be useful in isotropic simulations, in an anisotropic calculation with one fine direction, it may be advantageous to use a shorter timestep to update the links in that direction to ameliorate the typically larger forces that result from the shorter lattice spacing~\cite{Morrin:2006tf}. The anisotropy in step size requires a small amount of manual fine tuning, but should be similar to the
anisotropy in the lattice spacings. 

Our base integration scheme in this work is due to
Omelyan~\cite{Sexton:1992nu,Omelyan:2003,Takaishi:2005tz}; we
use the combined update operator
\begin{equation}
\mathcal{U}^1(\delta \tau) = \mathcal{U}_p(\lambda \delta \tau)
\mathcal{U}_q( \frac{1}{2} \delta \tau ) \mathcal{U}_{p}(1 - {2}
\lambda \delta \tau) \mathcal{U}_{q}( \frac{1}{2} \delta \tau )
\mathcal{U}_p( \lambda \delta \tau),
\end{equation}
which results in a scheme that is clearly reversible and is accurate to
$O(\delta \tau^3)$. The size of the leading error term can be further
minimized by tuning the parameter $\lambda$. In our work we used the
value of $\lambda$ from Ref.~\cite{Takaishi:2005tz} without any further
tuning, which promises an efficiency increase of approximately 50\%
over the simple leapfrog algorithm.

In Refs.~\cite{Sexton:1992nu,Weingarten:1980hx} it was shown that a
reversible, multi-level integration scheme can be constructed which
allows various pieces of the Hamiltonian to be integrated at different
timescales. Let us consider a Hamiltonian of the form
\begin{equation}
H(\pi, U)=\frac{1}{2} \pi^\dagger_{\mu}(x) \pi_{\mu}(x) + S_1(U) + S_2(U),
\end{equation}
where $S_1(U)$ and $S_2(U)$ are pieces of the action with corresponding
MD forces $F_1$ and $F_2$ respectively. One can then split the
integration into 2 timescales. One can integrate with respect to
action $S_1(U)$ using $\mathcal{U}^1(\delta \tau_1)$, where in the
component $\mathcal{U}_{p}(\delta \tau_1)$ we use only the force $F_{1}$. The
whole system can then be integrated with the update
\begin{equation}
\mathcal{U}^2(\delta \tau_2) = \mathcal{U}^\prime_p(\lambda \delta \tau_2)
\mathcal{U}^1( \frac{1}{2} \delta \tau_2 ) \mathcal{U}^\prime_{p}(1 - {2}
\lambda \delta \tau_2) \mathcal{U}^1( \frac{1}{2} \delta \tau_2 )
\mathcal{U}^\prime_p( \lambda \delta \tau_2),
\end{equation}
where in $\mathcal{U}^\prime_p$ we update the momenta using only $F_2$. Thus we end up with two characteristic integration timescales $\delta
\tau_1$ and $\delta \tau_2$. The scheme generalizes recursively to a
larger number of scales. A criterion for tuning the algorithm is to
arrange for terms in the action to be mapped to different timescales
so that on two timescales $i$ and $j$ we have $|| F_i || \delta
\tau_{i} \approx ||F_j|| \delta \tau_j$, as suggested in
Ref.~\cite{Hasenbusch:2002ai}.  We now proceed to outline how we split
our action.

We can write our gauge action schematically as
\begin{equation}
S = S_s(U) + S_t(U),
\end{equation}
where the term $S_s$ contains only loops with spatial gauge links, and
the $S_t$ term contains loops with spatial and temporal
links. While the term $S_s$ produces forces only in the spatial
directions, the $S_t$ term produces forces in both the spatial and the
temporal directions. In particular the spatial forces from $S_t$ are
larger in magnitude than the spatial forces from $S_s$ by roughly the
order of the anisotropy, and in turn, the temporal forces from $S_t$
are larger than the spatial forces from $S_t$. Our anisotropic
integration step size balances the spatial and temporal forces of the
$S_t$ term against each other. However, in order to balance the spatial
forces from $S_t$ and $S_s$ against each other, we integrate them on
separate time scales.

\subsection{Mass Preconditioning}
Following the work of~\cite{Hasenbusch:2002ai} our fermion determinant for the two flavor simulation can be written as
\begin{equation}
\det \left( Q^\dagger Q \right) = \frac{\det \left( Q^\dagger Q \right)}{
 \det \left( Q^\dagger_h Q_h \right)}  \det \left( Q_h^\dagger Q_h \right),
\end{equation}
where $Q$ is the fermion matrix with our desired fermion mass $m$ 
and $Q_h$ is the fermion matrix for which we choose a heavier
fermion mass $m_h$. After introducing pseudofermions the fermion action 
can be written as
\begin{equation}
S_{f} = S^{1}_{f} + S^2_{f},
\end{equation}
where
\begin{eqnarray}
S^1_{f} &=& \phi_1^{\dagger} Q_h \left( Q^\dagger Q \right)^{-1} Q^\dagger_h \phi_1 ,\\
S^2_{f} &=& \phi_2^{\dagger} \left( Q_h^\dagger Q_h \right)^{-1} \phi_2.
\end{eqnarray}
This trick introduces two main advantages: first, because $Q_h$ is
heavier than $Q$, inversion in $S^2_{f}$ will take fewer iterations
than solving with $Q$ directly, and forces
resulting from $Q_h$ will likewise be smaller than those that would
result from $Q$ allowing slightly longer time steps; second, as long
as $m$ is not very different from $m_h$, we have that to first order $
Q_h \left( Q^\dagger Q \right)^{-1} Q^\dagger_h \approx 1 + \Delta $
and that fluctuations with gauge fields will be to first order given
by $\frac{\delta \Delta}{\delta U}$. It should be clear, that as $m_h
\rightarrow m$ we have $\Delta \rightarrow 0$, and that the resulting
force $F \rightarrow 0$, in other words, that the magnitude of the
force from $S^1_f$ can be made small in a controlled manner. The
result is that while the inversions in $S^{1}_f$ are performed with
$Q$ and can be quite costly; by choosing $m_h$ appropriately the force
from $S^{1}_f$ can be reduced so that $S^{1}_f$ can be put on a long
time scale and evaluated relatively infrequently during an MD
trajectory. Some amount of effort is required to tune $m_h$ so that
the number of force evaluations from $S^1_f$ can be suitably reduced,
while at the same time keeping $m_h$ heavy enough, so that the force
evaluations and inversions from $S^2$ do not become overly expensive.

\subsection{Chronological Inversion Methods}
In order to further reduce our inversion costs, we employ chronological 
guesses in our MD. Before every new solve we produce a chronological
guess by employing the Minimal Residual Extrapolation method (MRE) of~\cite{Brower:1995vx}. This method works by using the last $n$ solution vectors, 
which are orthonormalized with respect to each other to create an $n$
dimensional basis. Let us denote these basis vectors $v_i$. The new 
initial guess $v_g$ is then constructed as
\begin{equation}
v_{g} = \sum_i a_i v_i \quad i=1 \hdots n,
\end{equation}
where $a_i$ are coefficients to be determined given the constraint
that the resulting $v_g$ minimize the functional minimized by the 
Conjugate Gradients process in the subspace spanned by $v_i$:
\begin{equation}
\Psi \left[ v_g \right] = v^\dagger_g Q^\dagger Q v_g - \chi^\dagger v_g - v_g^\dagger \chi,
\end{equation}
where $\chi$ is the right-hand side of the linear system for which the 
initial guess is being generated. Minimizing the functional $\Psi$ with respect to $a_i$ leads to the following set of linear equations for $a_i$:
\begin{equation}
\sum_{i=1}^{N} \left( v^{\dagger}_j Q^\dagger Q v_i \right) a_i = v_j^\dagger \chi.
\end{equation}

We emphasize that the use of chronological solution methods introduces
reversibility violations into the MD evolution,
and so the equations must be solved essentially exactly to avoid 
the reversibility violations from becoming large, and affecting
the detailed balance condition and thereby biasing the Monte Carlo
Markov process. To this end in our simulations we required a relative
stopping residuum of
\begin{equation}
r_{MD} = \frac{|| \chi - (Q^\dagger Q) \phi ||}{|| \chi ||} < 10^{-8}.
\end{equation}

\subsection{Summary}
In summary, our HMC algorithm uses a Hamiltonian composed of
the kinetic piece, the two gauge action pieces $S_s$ and $S_t$
and the two fermion action pieces $S^1_{f}$ and $S^2_{f}$. Our 
Molecular Dynamics evolution uses a 2nd Order Omelyan integrator split
over three timescales:
\begin{itemize}
\item 
Time scale 1 is the slowest, with time step $\delta \tau_1$, and is used to evolve the Hasenbusch ratio term with action $S^{1}_f$;
\item
Time scale 2 is faster, with time step $\delta \tau_2$, and it is used to evolve the 
mass preconditioned fermion term with action $S^{2}_f$ and the spatial gauge term with action $S_s$;
\item
Time scale 3 is the fastest, with time step $\delta \tau_3$, 
and it is used to evolve the temporal gauge term with action $S_t$.
\end{itemize}
Our overall MD trajectory length is set to be $\tau=1.0$.
In addition at all levels of the integrator, the spatial time step 
on that level is a factor of $\xi_{MD}=2.4$ larger than the temporal step.
Both fermionic terms use the MRE chronological guess technique with up
to $n=8$ previous solutions. These preconditioning masses $m_h$ and the 
concrete step sizes are summarized in Table \ref{tab:MDParams}.

The acceptance rates were typically between 60\% and 70\%.  The
simulations at mass $m=-0.4125$ made use of the QCDOC supercomputer
\cite{Chen:2000ej}, as well as BlueGene Teragrid Resource at San Diego
Supercomputer Center, while the entire $m=-0.4086$ dataset was generated
on Jaguar, a Cray XT3 resource at the National Center for Computational Science
(NCCS) at Oak Ridge National Laboratory through the INCITE'07 program
The HMC algorithm with the various improvements discussed in this section
is implemented and is freely available as part of the Chroma software system~\cite{Edwards:2004sx}.

\begin{table}
\begin{tabular}{|c|c|c|c|c|c|} 
\hline $\beta$ & $m_0$ & $m_h$ & $\delta \tau_1$ & $\frac{\delta \tau_2}{\delta \tau_1}$ & $\frac{ \delta \tau_3}{\delta \tau_2}$ \\
\hline
$5.5$ & $-0.4086$ & $-0.3700$ & ${1 \over 4}$ & ${1 \over 2}$ & ${1 \over 3}$ \\
$5.5$ & $-0.4125$ & $-0.3740$ & ${1 \over 4}$ & ${1 \over 4}$ & ${1 \over 3}$\\
\hline
\end{tabular}
\caption{The mass preconditioning masses $m_h$ and the time steps used
  in the Omelyan integration scheme in our simulations, for each
  target sea quark mass $m_0$. Except for time scale 1, the time step
  for each time scale is given relative to the previous
  one. Trajectories are of length $\tau=1$, with a step size
  anisotropy of $\xi_{MD}= 2.4$. The target solver residuum was
  $r=10^{-8}$ for both MD and Energy calculations and each fermionic
  term employed the MRE chronological guess method with up to the last
  8 previous vectors.\label{tab:MDParams}}
\end{table}

%% file: conventional2.tex
\section{Conventional Spectroscopy}\label{sec:conventional}

\subsection{Meson spectrum}\label{subsec:SimpleMeson}

\begin{figure}
\begin{center}
\begin{tabular}{cc}
\includegraphics[width=0.4\textwidth]{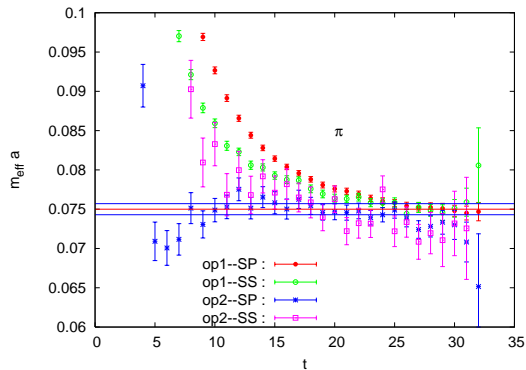}
\end{tabular}
\caption{Pion effective mass results for different smearings (S denotes smeared, P denotes point)
and operators (op1 denotes operators of Table~\ref{tab:meson_operator} and op2 denotes
operators including an extra factor of $\gamma^4$). Results are based
on 862 gauge configurations, quark mass parameter $m_0$ = -0.4125 
and a 24$^3\times$64 lattice.}
\label{fig:pion_meff}
\end{center}
\end{figure}

We use meson interpolating fields of the form $\bar{\psi}\Gamma\psi$ and 
$\bar{\psi}\Gamma\gamma^4 \psi$ 
that overlap with the physical states listed in
Table~\ref{tab:meson_operator}.  Correlation functions are calculated and we fit 
them with the analytical function, 
\begin{equation}
C(t) = A\Big( e^{-mt} + e^{-m(T -t)} \Big)  ,
\end{equation}
where $m$ is the mass parameter and $T$ is the time extent of the lattice.
Results for the mass parameters obtained from the fits of the correlations functions are summarized in
Tables~\ref{tab:meson_masses_416} and \ref{tab:meson_masses_578}. Comparisons of the fits for 
the $\pi$ meson with the effective mass are shown in
Figure~\ref{fig:pion_meff}
for the case of quark mass parameter $m_0 a_t$= -0.4125. 
Horizontal lines show the 
corresponding pion mass parameter of the fits and the error band.

\begin{table}
\begin{center}
\begin{tabular}{|c|c|c|}
\hline
$J^{PC}$ & $\Gamma$                 & $I=1$  \\
\hline
$0^{-+}$ & $\gamma_5$               & $\pi$  \\
$1^{--}$ & $\gamma_\mu$             & $\rho$ \\
$0^{++}$ & 1                        & $a_0$  \\
$1^{++}$ & $\gamma_\mu\gamma_5$     & $a_1$  \\
$1^{+-}$ & $\gamma_\mu\gamma_\nu$   & $b_1$  \\
\hline
\end{tabular}
\end{center}
\caption{\label{tab:meson_operator} Meson interpolating operators. 
The indicated charge-conjugation ($C$) quantum numbers apply 
only to particles with zero net flavor.}
\end{table}

\begin{table}
{\footnotesize	
\begin{tabular}{|cc|c|c|}
\hline
 Source &  Sink &  $m_\pi$ &  $m_\rho$    \\
\hline
 S & P$_1$ &   0.0754(5) &          \\
 S & S$_1$ &  0.0748(5) &  0.1430(7) \\
 S & P$_2$ &  0.0747(7)&   0.1438(10) \\
 S & S$_2$ &  0.0747(5) &  0.1427(8) \\
 S & P$_1$\&P$_2$ &0.0753(5)&0.1431(7) \\
 S & S$_1$\&S$_2$&0.0747(5)& 0.1427(8)\\
\hline
Result &   & 0.0750(7)  &  0.1431(8)\\
\hline
\end{tabular}
}
\caption{\label{tab:meson_masses_416}Meson masses (in temporal lattice units) for $N_f = 2$ with 
light quark mass $m_0 a_t = -.4125$ based on 862 configurations. 
Columns 1 and 2 
label the sources and sinks as smeared ($S$) or point ($P$) and 
the correlation functions based on the operators of Table~\ref{tab:meson_operator} (S$_1$ or P$_1$),
or based on including an extra factor $\gamma^4$ in the operators of Table~\ref{tab:meson_operator} 
(S$_2$ or P$_2$). Simultaneous fits are performed for two types of operators
in the results of rows 5 and 6.  Fits are performed in time windows of (26-32) with
$\chi^2$/DOF = 0.45(15) for the $\pi$ meson. }
\end{table}

\begin{table}
\hspace*{-1.3in}
\begin{center}
{\footnotesize
\begin{tabular}{|cc|c|c|}
\hline
Source & Sink & $m_{\pi}$ &   $m_\rho$ \\
\hline
S & P$_{1}$& 0.1088(7)& 0.1668(12)\\
S & S$_{1}$&  & 0.1652(13)\\
S & P$_{2}$& 0.1088(12)&0.1680(13)\\
S & P$_{1}$\&S$_{1}$ & & 0.1670(11)\\
S & P$_{1}$\&P$_{2}$ & 0.1088(7)& \\
\hline
Results & & 0.1088(8)&0.1668(16)\\
\hline
\end{tabular}}
\caption{\label{tab:meson_masses_578}Meson masses (in temporal lattice units) for $N_f = 2$ with 
quark mass $m_0 a_t = -.4086$ based on 363 configurations. 
Notation is the same as in Table~\ref{tab:meson_masses_416}.
Fits are performed in time windows of (26-32) with
$\chi^2$/DOF = 0.45(15) for the $\pi$ meson. }
\end{center}
\end{table}

\begin{figure}
\begin{tabular}{cc}
\includegraphics[width=0.4\textwidth,height=2.5in]{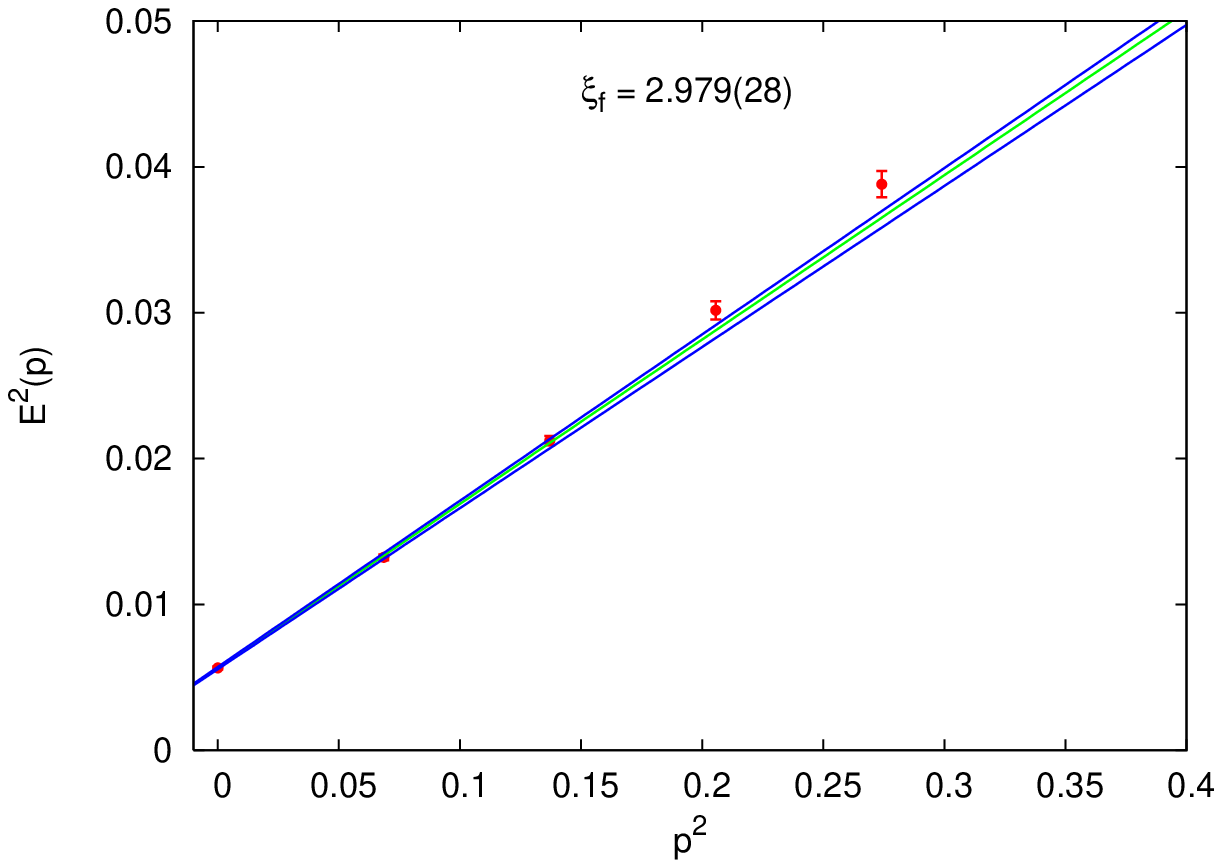} 
\includegraphics[width=0.4\textwidth,height=2.5in]{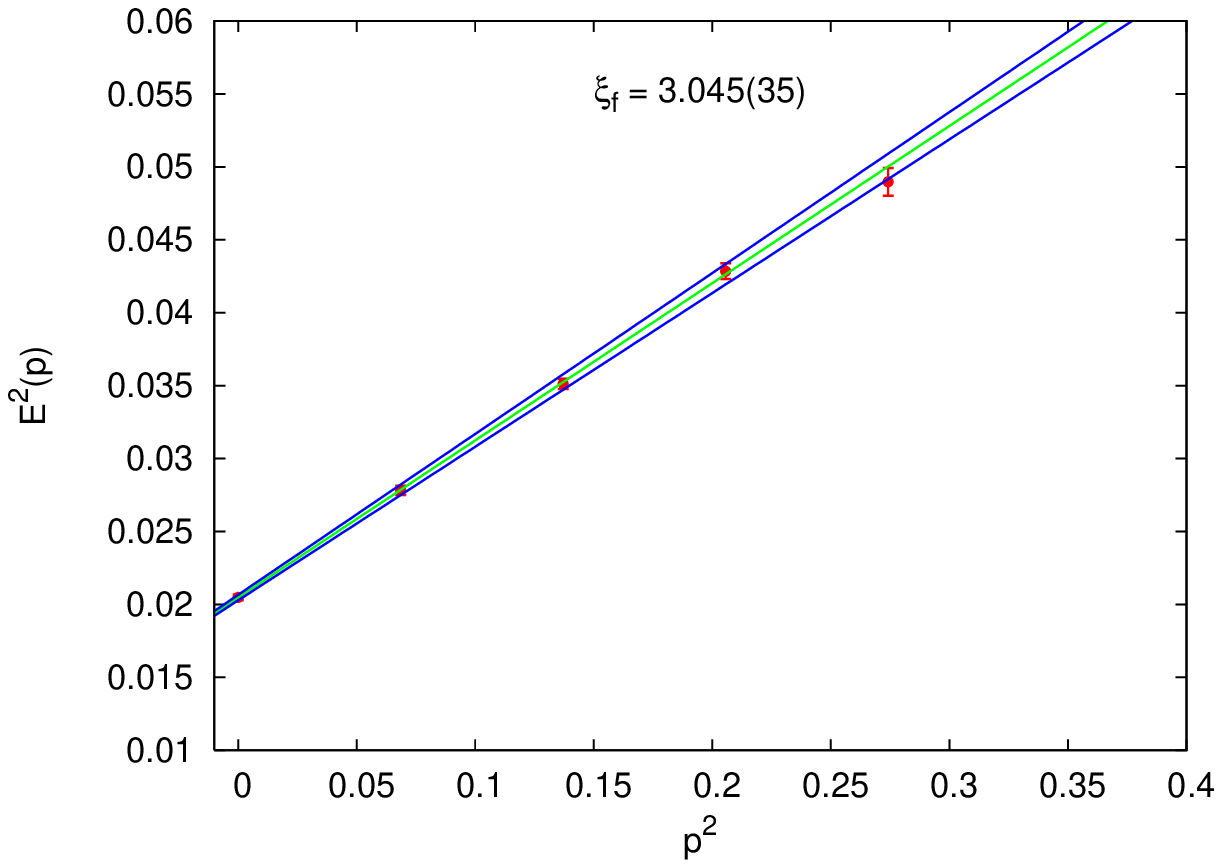}
\end{tabular}
\caption{ Dispersion relation for $\pi$ (left panel) and $\rho$ (right panel) mesons for quark mass parameter $m_0$ = -0.4125 and 24$^3\times$64 lattice. 
}
\label{fig:dispersion_pi_rho}
\end{figure}

The fermion anisotropy $\xi_f$ is determined through the conventional relativistic meson dispersion relation: 
\begin{equation}
E^2({\bf p}) = m^2 + \frac{{\bf p}^2}{\xi_f^2},
\label{eq:dispersion-rel}
\end{equation}
where the energy $E({\bf p})$ and the mass $m$ are in units of $a_t$, 
and ${\bf p} = \frac{2 \pi {\bf n}}{L_s}$ where $L_s$ is the spatial lattice size in units of $a_s$. 
From the two-point correlation functions we calculate the energy $E$ 
at the spatial momenta ${\bf p} = \frac{2\pi{\bf n}}{L_s}$ for 
${\bf n } = (0,0,0),(1,0,0),(1,1,0)$ and $(2,0,0)$ (averaged). 
The fitted jackknife energies are used in a linear fit 
of $E^2({\bf p})$ as a function of ${\bf p}^2$ as in Eq.~(\ref{eq:dispersion-rel}) 
in order to extract $\xi_f$. 
Figure~\ref{fig:dispersion_pi_rho} shows the dispersion relations
for $\pi$ and $\rho$ mesons. The fitted values of $\xi_f$
 are 2.979(28) (with first three momenta fit) for the $\pi$ meson and 3.045(35) 
(with first four momenta fit) for the rho meson. The central (green) line 
shows the fit and the blue bands show the errors. The desired 
fermion anisotropy matches the gauge anisotropy, $\xi_g$, which is 3 in our case.

%% file: scale.tex
\section{Scale Setting}\label{sec:scale}

In order to set the scale, the heavy-quark (static) potential $V(r)$
is calculated on a 16$^3\times$64 lattice. This is expected to have the form
\begin{eqnarray}
V(r) = C + \frac{\alpha}{r} + \sigma r, 
\label{eq:static_pot_th}
\end{eqnarray}
where $r$ is the separation between the static quarks. The scale implied by the heavy quark potential is often specified using the Sommer parameter $r_0$ which is defined by the condition
\begin{eqnarray}\label{eq:SQPr0}
- r^2 \frac{\partial V(r)}{\partial r} \Biggr|_{r=r_0} = 1.65.
\end{eqnarray}

On the lattice, we calculate Wilson loops which determine the static quark potential via
\begin{eqnarray}\label{eq:StaticQuarkPot}
W(r,t) &=& A e^{-V(r)t}.
\end{eqnarray}
In order to improve the signal and to extract the potential $V(r)$ from smaller time
separations, we smear the gauge links in the spatial directions using stout smearing
with parameters $n_{\rho}=16, n_{\rho}\rho=2.5$. We fit the Wilson loops as a
function of $t$ at each available $r$ to determine $V(r)$. We further fit $V(r)$ to
determine $C$, $\alpha$ and $\sigma$ according to Eq.~(\ref{eq:static_pot_th}) using
a standard jackknife procedure. Putting these parameters back into Eq.~(\ref{eq:SQPr0}),
we solve for $r_0/a_s$.

Finally, we relate this to the physical scale by using the value $r_0$= 0.462(11)(4) fm
from Refs.~\cite{Aubin:2004wf,Bernard:2006wx}
and set the scale $a_s$.  The results are summarized in
Table ~\ref{tab:Scale_set}.
\begin{table}[h]
\begin{tabular}{|c|c|c|c|c|c|c|c|}
\hline
$r_0$(fm)   &  $m_la_t$  &  $r_0/a_s$ &  $a_s$ (fm)   &  $a_t^{-1}$(MeV)& $m_{\pi}a_t$ & $m_{\pi}(MeV) $ &  $\xi_0$ \\
\hline
0.462(11)(4)& -0.4086 &  4.10(8)   &  0.113(7)&  5310(265)      &  0.1088(37)  &  578(29)&  2.38 \\
0.462(11)(4)& -0.4125 &  4.26(12)  &  0.108(7)&  5556(333)      &  0.0750(24)  &  416(36)&   2.38 \\
\hline
\end{tabular}
\caption{ \label{tab:Scale_set} The value of the Sommer parameter, $r_0$,
is listed in column 1 and the ratio $r_0/a_s$ for each quark
mass on our 16$^3\times$64 lattices is listed in column 3.  The scale obtained from $r_0/(r_0/a_s)$ is listed in column 4.
Using the renormalized anisotropy $\xi =$ 3, we find the temporal spacing $a_t^{-1}$
as given in column 5.  The pion mass in lattice units is given in column 6 and
in MeV units in column 7, while the bare anisotropy $\xi_0$ is given in column 8.}
\end{table}

%% file: operators2.tex
\section{Baryon Operators} \label{sec:baryon-ops}

The use of operators whose temporal correlation functions attain their
asymptotic form as quickly as possible is crucial for reliably
extracting excited hadron masses.  An important ingredient in constructing
such hadron operators is the use of smeared fields.  Operators constructed
from smeared fields have dramatically reduced mixings with the high frequency
modes of the theory.  Both link-smearing and quark-field smearing are
necessary.  Since excited hadrons are expected to be large objects, 
the use of spatially extended operators is another key ingredient in
the operator design and implementation.  

\subsection{Smearing} 
Spatial links can be smeared using the stout-link procedure described in
Ref.~\cite{Morningstar:2003gk}. The stout-link smearing scheme is analytic, efficient,
and produces smeared links that automatically are elements of $SU(3)$ 
without the need for a projection back into $SU(3)$.  
Note that only spatial staples are used in the link smoothing; no temporal
staples are used, and the temporal link variables are not smeared.
The smeared quark fields can be defined by
\begin{equation}
\widetilde{\psi}(x) = \left(1+\frac{\sigma_s^2}{4n_\sigma}\ \widetilde{\Delta}
 \right)^{n_\sigma}\psi(x),
\end{equation}
where $\sigma_s$ and $n_\sigma$ are tunable parameters ($n_\sigma$ is 
a positive integer) and the three-dimensional covariant Laplacian
operators are defined in terms of the smeared link variables 
$\widetilde{U}_j(x)$ as follows: 
\begin{eqnarray}
 \widetilde{\Delta} O(x) &=& \!\!\!\sum_{k=\pm 1,\pm2,\pm3} \biggl(
  \widetilde{U}_k(x)O(x\!+\!\hat{k})-O(x) \biggr), 
\end{eqnarray}
where $O(x)$ is an operator defined at lattice site
$x$ with appropriate color structure, and noting that
$\widetilde{U}_{-k}(x)=\widetilde{U}_k^\dagger(x\!-\!\hat{k})$.  
The smeared fields $\widetilde{\psi}$ and $\widetilde{\overline{\psi}}$ are 
Grassmann-valued; in particular, these fields anticommute in the same way
that the original fields do, and the square of each smeared field vanishes.

\subsection{Group theory} 
Hadron states are identified by their momentum $\pvec$, intrinsic
spin $J$, projection $\lambda$ of this spin onto some axis,
parity $P=\pm 1$, and quark flavor content (isospin, strangeness, {\it etc.}).
Some mesons also include $G$-parity as an identifying quantum number.
If one is interested only in the masses of these states, one can restrict
attention to the $\pvec=\bm{0}$ sector, so operators must be invariant 
under all spatial translations allowed on a cubic lattice.  The little 
group of all symmetry transformations on a cubic lattice which leave 
$\pvec=\bm{0}$ invariant is the octahedral point group $O_h$, so operators
may be classified using the irreducible representations (irreps) of $O_h$.
For mesons, there are ten irreducible representations $A_{1g}, A_{2g}, 
E_g, T_{1g}, T_{2g}, A_{1u}, A_{2u}, E_u, T_{1u}, T_{2u}.$  The 
representations with a subscript $g (u)$ are even (odd) under parity.
The $A$ irreps are one dimensional, the $E$ irreps are two dimensional,
and the $T$ irreps are three-dimensional.
The $A_1$ irreps contain the $J=0,4,6,8,\dots$ states, the $A_2$ irreps
contain the $J=3,6,7,9,\dots$ states, the $E$ irreps contain the
$J=2,4,5,6,7,\dots$ states, the $T_1$ irreps contain the
spin $J=1,3,4,5,\dots$ mesons, and the $T_2$ irreps contain the spin 
$J=2,3,4,5,\dots$ states.  For baryons, there are four two-dimensional irreps 
$G_{1g}, G_{1u}, G_{2g}$, $G_{2u}$ and two four-dimensional representations 
$H_g$ and $H_u$.  The $G_1$ irrep contains the
$J=\frac{1}{2},\frac{7}{2},\frac{9}{2},\frac{11}{2},\dots$ states,
the $H$ irrep contains the $J=\frac{3}{2},\frac{5}{2},\frac{7}{2},
\frac{9}{2},\dots$ states, and the $G_2$ irrep contains the 
$J=\frac{5}{2},\frac{7}{2},\frac{11}{2},\dots$ states.
The continuum-limit spins $J$ of our states must be deduced by examining
degeneracy patterns across the different $O_h$ irreps.

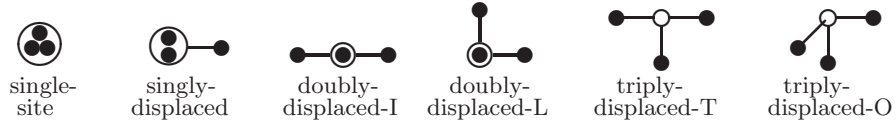
\begin{figure}[t]
\centerline{
\raisebox{0mm}{\setlength{\unitlength}{1mm}
\thicklines
\begin{picture}(16,10)
\put(8,6.5){\circle{6}}
\put(7,6){\circle*{2}}
\put(9,6){\circle*{2}}
\put(8,8){\circle*{2}}
\put(4,0){single-}
\put(5,-3){site}
\end{picture}}
\raisebox{0mm}{\setlength{\unitlength}{1mm}
\thicklines
\begin{picture}(16,10)
\put(7,6.2){\circle{5}}
\put(7,5){\circle*{2}}
\put(7,7.3){\circle*{2}}
\put(14,6){\circle*{2}}
\put(9.5,6){\line(1,0){4}}
\put(4,0){singly-}
\put(2,-3){displaced}
\end{picture}}
\raisebox{0mm}{\setlength{\unitlength}{1mm}
\thicklines
\begin{picture}(20,8)
\put(12,5){\circle{3}}
\put(12,5){\circle*{2}}
\put(6,5){\circle*{2}}
\put(18,5){\circle*{2}}
\put(6,5){\line(1,0){4.2}}
\put(18,5){\line(-1,0){4.2}}
\put(6,0){doubly-}
\put(4,-3){displaced-I}
\end{picture}}
\raisebox{0mm}{\setlength{\unitlength}{1mm}
\thicklines
\begin{picture}(20,13)
\put(8,5){\circle{3}}
\put(8,5){\circle*{2}}
\put(8,11){\circle*{2}}
\put(14,5){\circle*{2}}
\put(14,5){\line(-1,0){4.2}}
\put(8,11){\line(0,-1){4.2}}
\put(4,0){doubly-}
\put(1,-3){displaced-L}
\end{picture}}
\raisebox{0mm}{\setlength{\unitlength}{1mm}
\thicklines
\begin{picture}(20,12)
\put(10,10){\circle{2}}
\put(4,10){\circle*{2}}
\put(16,10){\circle*{2}}
\put(10,4){\circle*{2}}
\put(4,10){\line(1,0){5}}
\put(16,10){\line(-1,0){5}}
\put(10,4){\line(0,1){5}}
\put(4,0){triply-}
\put(1,-3){displaced-T}
\end{picture}}
\raisebox{0mm}{\setlength{\unitlength}{1mm}
\thicklines
\begin{picture}(20,12)
\put(10,10){\circle{2}}
\put(6,6){\circle*{2}}
\put(16,10){\circle*{2}}
\put(10,4){\circle*{2}}
\put(6,6){\line(1,1){3.6}}
\put(16,10){\line(-1,0){5}}
\put(10,4){\line(0,1){5}}
\put(4,0){triply-}
\put(2,-3){displaced-O}
\end{picture}}  }
\vspace*{8pt}
\caption{The spatial arrangements of the extended three-quark baryon
operators. Smeared quark-fields are
shown by solid circles, line segments indicate
gauge-covariant displacements, and each hollow circle indicates the location
of a Levi-Civita color coupling.  For simplicity, all displacements
have the same length in an operator.  Results presented here used
displacement lengths of $3a_s$ ($\sim 0.3 \mbox{ fm}$).
\label{fig:operators}}
\end{figure}

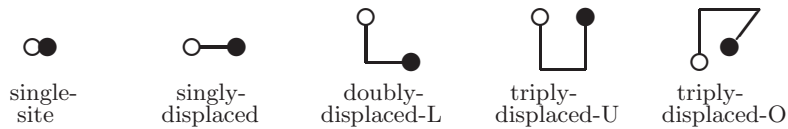
\begin{figure}[b]
\centerline{
\raisebox{0mm}{\setlength{\unitlength}{1mm}
\thicklines
\begin{picture}(20,12)
\put(7,7){\circle{2}}
\put(9,7){\circle*{2.5}}
\put(4,0){single-}
\put(5,-3){site}
\end{picture}}
\raisebox{0mm}{\setlength{\unitlength}{1mm}
\thicklines
\begin{picture}(20,12)
\put(6,7){\circle{2}}
\put(12,7){\circle*{2.5}}
\put(7,7){\line(1,0){4}}
\put(4,0){singly-}
\put(2,-3){displaced}
\end{picture}} 
\raisebox{0mm}{\setlength{\unitlength}{1mm}
\thicklines
\begin{picture}(20,12)
\put(7,11){\circle{2}}
\put(13,5){\circle*{2.5}}
\put(12,5){\line(-1,0){5}}
\put(7,10){\line(0,-1){5}}
\put(4,0){doubly-}
\put(1,-3){displaced-L}
\end{picture}}
\raisebox{0mm}{\setlength{\unitlength}{1mm}
\thicklines
\begin{picture}(20,12)
\put(8,11){\circle{2}}
\put(14,11){\circle*{2.5}}
\put(8,4){\line(1,0){6}}
\put(14,4){\line(0,1){6}}
\put(8,4){\line(0,1){6}}
\put(4,0){triply-}
\put(2,-3){displaced-U}
\end{picture}}
\raisebox{0mm}{\setlength{\unitlength}{1mm}
\thicklines
\begin{picture}(20,15)
\put(7,5){\circle{2}}
\put(11,7){\circle*{2.5}}
\put(7,12){\line(1,0){8}}
\put(7,6){\line(0,1){6}}
\put(15,12){\line(-3,-4){3.0}}
\put(4,0){triply-}
\put(2,-3){displaced-O}
\end{picture}} }
\caption[captab]{The spatial arrangements of the quark-antiquark meson operators.
In the illustrations, the smeared quarks fields are depicted by solid circles, 
each hollow circle indicates a smeared ``barred'' antiquark field, and the 
solid line segments indicate covariant displacements.
\label{fig:mesops}}
\end{figure}

\subsection{Operator construction and pruning}
Our operators are constructed in a three-stage approach\cite{Basak:2005aq}.
First, basic building blocks are chosen.  These are taken to be smeared 
covariantly-displaced quark fields
\begin{equation}
\bigl(\widetilde{D}^{(p)}_j\ \widetilde{\psi}\bigr)^A_{a\alpha},
 \hspace{0.3in}  \bigl(\widetilde{\overline{\psi}}  
   \ \widetilde{D}^{(p)\dagger}_j \bigr)^A_{a\alpha},
 \qquad -3\leq j\leq 3,
\label{eq:blocks}
\end{equation}
where $A$ is a flavor index, $a$ is a color index, $\alpha$ is a
Dirac spin index, and the $p$-link gauge-covariant displacement
operator in the $j$-th direction is defined by
\begin{equation}
 \widetilde{D}_j^{(p)}(x,x^\prime) =
 \widetilde{U}_j(x)\ \widetilde{U}_j(x\!+\!\hat{j})\dots 
   \widetilde{U}_j(x\!+\!(p\!-\!1)\hat{j})\delta_{x^\prime,x+p\hat{j}},
\qquad \widetilde{D}_0^{(p)}(x,x^\prime) = \delta_{xx^\prime},
\end{equation}
for $j=\pm 1,\pm 2,\pm 3$ and $p\geq 1$, and where $j=0$ defines
a zero-displacement operator to indicate no displacement. 
Next, {\em elemental} operators $B^{F}_i(t,\bm{x})$ are devised
having the appropriate flavor structure characterized by isospin, strangeness,
{\it etc.}, and color structure constrained by gauge invariance.
For zero momentum states, translational invariance is imposed:
$B^{F}_i(t)=\sum_{{\bm{x}}}  B^{F}_i(t,\xvec).$
Finally, group-theoretical projections are applied to
obtain operators which transform irreducibly under 
all lattice rotation and reflection symmetries:
\begin{equation}
  {\cal B}_{i}^{\Lambda\lambda F}(t)
 = \frac{d_\Lambda}{g_{O_h^D}}\sum_{R\in O_h^D} 
  \Gamma^{(\Lambda)}_{\lambda\lambda}(R)\ U_R\ B^F_i(t)\ U_R^\dagger,
\label{eq:project}
\end{equation}
where $O_h^D$ is the double group of $O_h$, $R$ denotes an element of $O_h^D$,
$g_{O_h^D}$ is the number of elements in $O_h^D$, and $d_\Lambda$ is the
dimension of the $\Lambda$ irreducible representation.  Projections onto
both the single-valued and double-valued irreps of $O_h$ require using the 
double group $O_h^D$ in Eq.~(\ref{eq:project}). 
Given $M_B$ elemental $B^F_i$ operators, many of the projections in 
Eq.~(\ref{eq:project}) vanish or lead to linearly-dependent operators,
so one must then choose suitable linear combinations of the projected operators
to obtain a final set of independent baryon operators.
Thus, in each symmetry channel, one ends up with a set of $r$ operators 
given in terms of a linear superposition of the $M_B$ elemental operators.
The different spatial configurations (see Fig.~\ref{fig:operators} for
the baryon configurations and Fig.~\ref{fig:mesops} for the meson
configurations) yield operators which effectively
build up the necessary orbital and radial structures of the hadron
excitations.  The design of these operators is such that a large number
of them can be evaluated very efficiently, and components in their
construction can be used for both meson, baryon, and multi-hadron
computations.

\begin{figure}[t]
\begin{center}
\includegraphics[width=4.0in, bb=0 40 567 559]{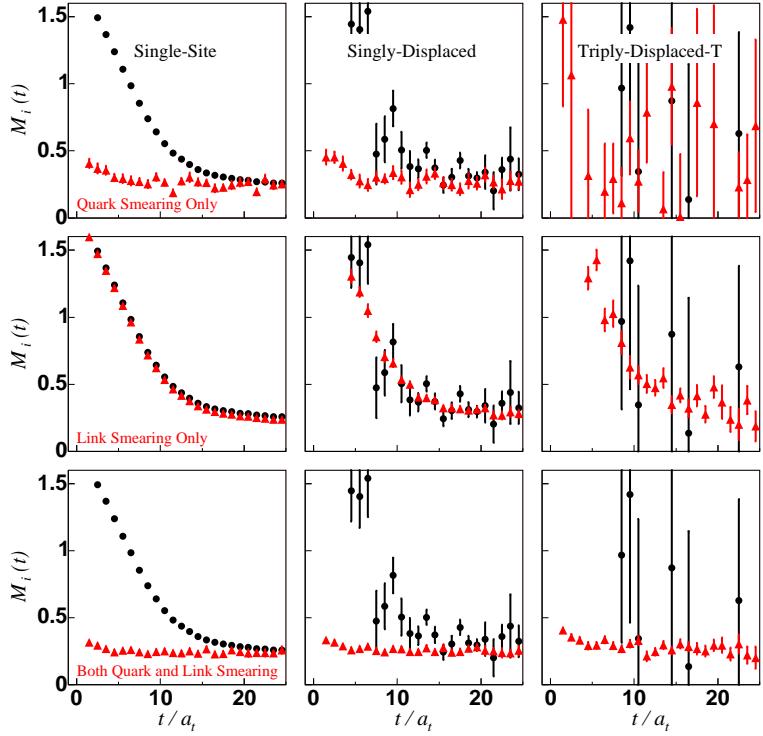}
\end{center}
\caption{Effective masses $M(t)$ for unsmeared (black circles) and smeared 
(red triangles) operators $O_{SS},\ O_{SD},\ O_{TDT}$, which are
representative single-site, singly-displaced, and triply-displaced-T 
nucleon operators, respectively.
Top row: only quark-field smearing $n_\sigma=32,\ \sigma_s=4.0$ is used. Middle row: 
only link-variable smearing $n_\rho=16,\ n_\rho\rho=2.5$ is applied.  
Bottom row: both quark and link smearing $n_\sigma=32,\ \sigma_s=4.0, 
\ n_\rho=16,\ n_\rho\rho=2.5$ are used, dramatically improving the signal for all
three operators. Results are based on 50 quenched configurations on a 
$12^3\times 48$ anisotropic lattice using the Wilson action with $a_s \sim 0.1$ fm,
$a_s/a_t \sim 3.0$.\label{fig:smearing}}
\end{figure}

Finding appropriate smearing parameters is a crucial initial part 
of any hadron spectrum calculation.  Fig.~\ref{fig:smearing} demonstrates 
that {\em both} quark-field and link-field smearing are needed in order for
spatially-extended baryon operators to be useful\cite{Basak:2005gi}.
It is important to use the smeared links when smearing the quark field.
Link smearing dramatically reduces the statistical errors in the
correlators of the displaced operators, while quark-field smearing
dramatically reduces the excited-state contamination.  In 
this study, the quark field is Gaussian smeared with $\sigma = 3.0$ using 
32 iterations, and the link field is stout-smeared with 
$n_{\rho}=16, n_\rho \rho= 2.5$. 

Our approach to designing hadron and multi-hadron interpolating
fields leads to a very large number of operators.  It is not
feasible to do spectrum computations using all of the operators 
so designed; for example, in the $G_{1g}$ symmetry channel for nucleons,
the above procedure leads to 179 operators.  It is necessary to \textit{prune}
down the number of operators.  After much exploratory testing and trials,
we found that a procedure that keeps a variety of operators while minimizing
the effects of noise works best for facilitating the extraction of several
excited states.  Some operators are intrinsically noisy and must be removed.
In addition, a set of operators, each with little intrinsic noise, can allow
noise to creep in if they are not sufficiently independent of one another.

The following procedure was used. (1) First, operators with excessive 
intrinsic noise were removed.  This was done by examining the diagonal 
elements of the correlation matrix and discarding
those operators whose self-correlators had relative errors above some
threshold for a range of temporal separations.  A low-statistics Monte Carlo
computation on a reasonably small lattice was used to accomplish this.
(2) Second, pruning within operator types (single-site, singly-displaced, 
\textit{etc.}) was done based on the condition number of the submatrices
\[
   \widehat{C}_{ij}(t) = \frac{C_{ij}(t)}{\sqrt{C_{ii}(t)C_{jj}(t)}},
\qquad t=a_t.
\]
The condition number was taken to be the ratio of the largest eigenvalue
over the smallest eigenvalue.  A value near unity is ideal.
For each operator type, the set of about six operators which yielded
the lowest condition number of the above submatrix was retained.
(3) Lastly, pruning across all operator types was done based again on 
the condition number of the remaining submatrix as defined above.  In this
last step, the goal was to choose about 16 operators, keeping two or three
of each type, such that a condition number reasonably close to unity was
obtained.  As long as a good variety of operators was retained, the 
resulting spectrum seemed to be fairly independent of the exact choice of
operators at this stage.  Eigenvectors from a variational study of the 
operators could also be used to fine tune the choice of operators.

%% file: results4.mod.tex
\section{I = $\frac{1}{2}$ baryon spectra }
\label{sec:results}

Each of our unbarred (barred) baryon operators annihilates 
(creates) a baryon and creates (annihilates) an antibaryon.  
This causes correlation functions to have
a baryon state propagating forward in time and an antibaryon state
propagating backward in time.  Because fermions and antifermions
 have opposite intrinsic parities, the backward
in time signal corresponds to states with parity opposite to that of the
 states propagating forward in time.  Because of PCT symmetry coupled with the use of
antiperiodic boundary conditions, correlation functions obey the rule
\begin{equation}
 C^{(\Lambda)}_{kk'}(t) = C^{(\Lambda_c)}_{kk'}(T-t)^*,
\end{equation}
where $\Lambda_c$ is the opposite parity partner of irrep $\Lambda$.  This allows us to increase
 our statistics by ``folding'' the correlation functions:
\begin{equation}
 C^{(\Lambda)}_{kk'}(t)\to \frac{1}{2}\left( C^{(\Lambda)}_{kk'}(t) + C^{(\Lambda_c)}_{kk'}(T-t)^*\right).
\end{equation}
Generally the separation of the two time slices involved is sufficient 
to provide independent samples of the gauge configurations. 

   We choose phases of our baryon operators such that the matrices of
correlation functions are real.  We also average the matrix with its transpose in order to guarantee that the 
matrices are symmetric. This helps to clean up the signals by
reducing the errors. 

\subsection{The Variational Method}
 We calculated 16$\times$16 matrices
of correlation function in each irrep of the octahedral group:
$\Lambda=\{ G_{1g}, G_{2g}, H_g, G_{1u}, G_{2u}, H_u\}$. 
The variational method was used to help extract the excited 
spectrum from the matrices of 
correlation functions, which involved numerically solving the generalized eigenvalue problem
\begin{equation}
 C_{kk'}^{(\Lambda)}(t)v_{k'}^{(n)}(t,t_0) = \lambda^{(\Lambda)}_n(t,t_0)C_{kk'}^{(\Lambda)}(t_0) v_{k'}^{(n)}(t,t_0),
\end{equation}
where $n$ labels the eigenstates.  
Degeneracies and numerical uncertainties can cause variances of the 
eigenvectors at different times.  We studied two methods 
for extracting the spectrum: 1) a fixed-eigenvector method and 
2) a principal-correlator method where the diagonalizations were performed 
on each time step.

 The fixed-eigenvector method  
involved solving the eigenvalue problem on a single time slice $t=t^*$
using a fixed value of $t_0$. These
eigenvectors are normalized with respect to
$C_{kk'}^{(\Lambda)}(t_0)$ such that
\begin{equation}
 v_k^{(n)\dag}(t^*,t_0)C_{kk'}^{(\Lambda)}(t_0) v_{k'}^{(n)}(t^*,t_0) =
 \delta_{kk'}.
\label{eq:evec_norm}
\end{equation}
For each time slice and each configuration, the matrix
of correlations functions was rotated to this
basis of vectors using
\begin{equation}
 \tilde C^{(\Lambda)}_{kk'}(t) = V_{kn}^\dag(t^*,t_0) C^{(\Lambda)}_{nn'}(t) V_{n'k'}(t^*,t_0),
\label{eq:rot_corr_mat}
\end{equation}
where $V_{n'k'}(t^*,t_0)$ is the matrix whose columns are the
eigenvectors at time $t^*$.  Because of Eq.~(\ref{eq:evec_norm}), 
the rotated matrices of correlation functions are equal to the
identity matrix at time $t_0$.

  The diagonal elements of the rotated correlation matrix are related to the energies by
\begin{equation}
 \tilde C_{kk}^{(\Lambda)}(t)\simeq e^{-E_k(t-t_0)} + \sum_{n \neq k} \alpha_n e^{-E_n(t-t_0)}
+ {\cal O}(e^{-(E_{N+1}-E_k)t}),
\label{eq:decay_energy}
\end{equation}
where the sum over terms involving $\alpha_n$ vanishes at time $t^*$, 
but can contribute away from $t^*$. The term involving the first omitted energy, $E_{N+1}$, has been derived by Blossier et al. \cite{Blossier:2008tx}.  
We extract the low lying energies by performing fully correlated $\chi^2$-minimization fits, modelling the $k^{th}$ diagonal element of the rotated correlator matrix as
\begin{equation}
\tilde C^{fit}_{kk}(t)=  (1 - A) e^{-E_k(t - t_0)} + Ae^{-E'(t-t_0)},\label{eq:fit_doub_exp}
\end{equation}
where $E_k$ is the energy of the $k^{th}$ state.  The second exponential captures the contribution of the higher energy states and allows us to fit the correlators to early time slices.  The choice of coefficents in front of each exponential enforces $\tilde C^{fit}_{kk}(t_0) = 1$, as guaranteed by Eq.~(\ref{eq:evec_norm}).  We assume that the $\alpha_n$ in Eq. \ref{eq:decay_energy} are negligible in the time range over which we perform the fit.

We optimized our choice of $t_0$ and $t^*$ using a method adapted from that in Ref.~\cite{Dudek:2007wv}.  In that work, the optimal choice of $t_0$ was determined for the extraction of the charmonium spectrum using a principal-correlator analysis.  This optimal choice balanced the need for the contributions of higher energy states to have decayed away (suggesting larger values for $t_0$) and for the correlator to have a low level of noise (suggesting smaller values for $t_0$).  The energies in the low lying spectrum were extracted by fitting the principal correlators for various values of $t_0$.  For each value of $t_0$, the correlator was  reconstructed from these fit energies and the eigenvectors using the spectral decomposition of the correlator matrix
\begin{equation}
C_{ij}(t) =  \langle {\cal O}_i(t) {\cal O}_j(0)\rangle = \sum_\alpha
\frac{Z_i^{\alpha*} Z_j^\alpha}{2 m_\alpha} e^{-m_\alpha t}. \label{spec}
\end{equation}
The overlap factors $Z^\alpha_i = \langle 0 | {\cal O}_i |
\alpha \rangle$ are related to the eigenvectors of the correlator by
\begin{equation}
Z^\alpha_i = (V^{-1} )^\alpha_i  \sqrt{2 m_\alpha} e^{m_\alpha t_0 / 2}. 
\end{equation}

A $\chi^2$-like quantity was defined to measure how well the reconstructed correlator described the original correlator matrix:
\begin{equation}
\chi^2 = \frac{1}{\tfrac{1}{2} N (N +1 ) (t_{\mathrm{max}} - t_0) - \tfrac{1}{2}N (N +3 )
}\sum_{i, j \geq i} \sum_{t, t'=t_0+1}^{t_{\mathrm{max}}} (C_{ij}(t) -
C^{\mathrm{rec.}}_{ij}(t)) \mathbb{C}_{ij}^{-1}(t, t')  (C_{ij}(t') - C^{\mathrm{rec.}}_{ij}(t')),\label{eq:chisq_corr}
\end{equation}
where $\mathbb{C}_{ij}^{-1}(t, t')$ is the correlation matrix for the correlator $C_{ij}$.  
Although the principal-correlator method actually yields time dependent overlap factors $Z(t)$ 
(because the correlator matrix is diagonalized on all time slices),  it was observed that the $Z(t)$ were reasonably 
constant and the reconstruction was done using a single $Z(t_Z)$ chosen at a 
time such that $\chi^2$ was minimized.  For $t_Z> t_0$, the variation in $\chi^2$ as 
a function of $t_Z$ was minimal.

In this work, we adapt this technique for the fixed eigenvector method, finding optimal values for $t_0$ and $t^*$.  We extract the $16$ lowest energies in the spectrum by fitting the diagonal elements of the rotated correlator matrix, Eq.~(\ref{eq:rot_corr_mat}), obtained using a range of values for $t_0$ and $t^*$.  Reconstructing the correlator from these masses and the $Z$ factors at $t^*$, we choose the $t_0$ and $t^*$ which minimize the $\chi^2$.

To correctly extract the energy spectrum, it is also crucial to select an appropriate range of time slices on which to fit the correlator.  In particular, we would like to avoid time slices where the opposite-parity backward-propagating state contributes to the correlator. For mesons, where the forward and backward-propagating states have the same parity, the
 variational method simultaneously diagonalizes the forward and backward-propagating parts
  of a meson correlation function.  This is not the case for baryons where
  the forward and backward-propagating states have opposite parities and different energies. 
    The forward-in-time signals dominate at small values of time but they decay exponentially
and the backward-propagating signals
    can become significant after some threshold value of time.  We were  able to extract the energies of the states 
by fitting the diagonal correlation
  functions using Eq.~(\ref{eq:fit_doub_exp}) without significant
interference from the backward propagating signal for all channels except $G_{1u}$ at $m_\pi=416$ MeV.
      In this channel, the backward propagating signal is dominated by the $G_{1g}$ ground state,
       which is the lowest energy state in the spectrum.  For our lattice at the lower pion mass, the backward-propagating
       $G_{1g}$ signal decayed slowly enough and the
       temporal extent was small enough (due to the anisotropy) that the $G_{1u}$ signals had significant backward contamination even at small time slices.
To extract the $G_{1u}$ energy levels using the fixed-eigenvector 
method, we include the backward propagating state in the fit and constrain its energy by 
fitting simultaneously the $G_{1g}$ ground state:
\begin{eqnarray}
C^{fit,G_{1u}}_k &=& (1-A-B)e^{-E^{G_{1u}}_k(t-t_0)} + Ae^{-E'(t-t_0)} + Be^{E^{G_{1g}}_0(t-t_0)},  \label{eq:fit_G1u}\\
C^{fit,G_{1g}}_0 &=& (1-D)  e^{-E^{G_{1g}}_0(t-t_0)} + De^{-E''(t-t_0)}.
\label{eq:fit_G1g}
\end{eqnarray}
Due to the increased noise in the excited states, the minimizer was unable to find a minimum in the $\chi^2$ for these simultaneous fits for $k\geq2$.  We were able to successfully fit these states by modeling the forward propagating state as single exponential and fitting only on later time slices (where the higher energy states had completely decayed).

The fit ranges were optimized such that the $\chi^2$ was minimized.  To visually confirm the sensibility of the fit parameters, we look at plots of  
\begin{equation}
\tilde C^{\Lambda}_{kk}e^{E_k (t-t_0)},
\label{eq:corr}
\end{equation}
versus time.
If Eq.~(\ref{eq:fit_doub_exp}) correctly models the correlator, then the plot should plateau to $(1-A)$ and we confirm that the plateau is consistent with the value of $A$ determined from the fit.  For the $G_{1u}$ channel at $m_\pi=416$ MeV, we first subtract off the backward exponential and compare the plateau with $(1-A-B)$ as in Eq.~(\ref{eq:fit_G1u}).
 Finally, we confirm that the fit parameters are stable under small variations in the fit range.  We estimate the uncertainty in the fit energy through a jackknife analysis.  We fit each member of a jackknife ensemble to obtain an ensemble of energies and report the average energy and the jackknife error.
 
The presence of the backward-propagating state in the $G_{1u}$
channel caused numerical instabilities in the eigenvectors of the 
principal-correlator method.
In order to remove
the cause of the problem, we tested a
method based on filtering out the backward signal prior to
diagonalization.  In a time interval where the backward signal is
simply the ground state of the opposite parity channel with energy
 $E^{\Lambda_c}_0$, the matrix of correlation functions can be modeled as
a forward part plus the single backward state,
\begin{equation}
 C_{kk'}^{(\Lambda)}(t)=\sum_{n} A_{k k'}^{(n)} e^{-E^\Lambda_n(t-t_0)} + B_{k k'} e^{-E_0^{\Lambda_c}(T-t_0)}.
\end{equation}
We define the filtered correlator as
\begin{eqnarray}
 C_{filt,kk'}^{(\Lambda)} &=& C_{kk'}^{(\Lambda)}(t) - C_{kk'}^{(\Lambda)}(t_1) +(1-e^{-E^{\Lambda_c}}_0) \sum_{j=t+1}^{t_1} C_{kk'}^{(\Lambda)}(j) ,
\end{eqnarray}
and find that it can be modeled as
\begin{eqnarray}
C_{filt,kk'}^{(\Lambda)}&=& \sum_n
\tilde{A}^{(n)}_{kk'}\left(e^{-E^{\Lambda}_n(t-t_0)}-
e^{-E^{\Lambda}_n(t_1-t_0)}\right), \nonumber \\
\tilde{A}^{(n)}_{kk'} &=&  A_{k k'}^{(n)} \left[1+\frac{1-e^{-E^{\Lambda_c}_0}}{e^{E^\Lambda_n}-1}\right],
\end{eqnarray}
where $t_1$ is a time where the backward signal is, in fact,
described by single exponential. The backward-in-time signal for 
energy $E^{\Lambda_c}_0$ is reduced to the level of errors 
and the filtered correlators consist of the renormalized forward 
signal minus a constant term. The diagonalization of
the filtered correlators using the principal-correlator method produced
stable eigenvectors and the energies of the states could be
extracted by fitting the principal correlation functions to a
single exponential decay with a constant term.  However, this method did
not produce any significant improvement over the results from the
fixed-eigenvector method. We point out that the filtering is 
necessary in order to extract the $G_{1u}$ excited spectrum from our lattices using the principal-correlator
method.
\subsection{Results}
\begin{figure}
\begin{center}
\begin{tabular}{lc}
\hspace{-1in}
\includegraphics[width=0.4\textwidth,clip=true,height=0.5\textwidth]{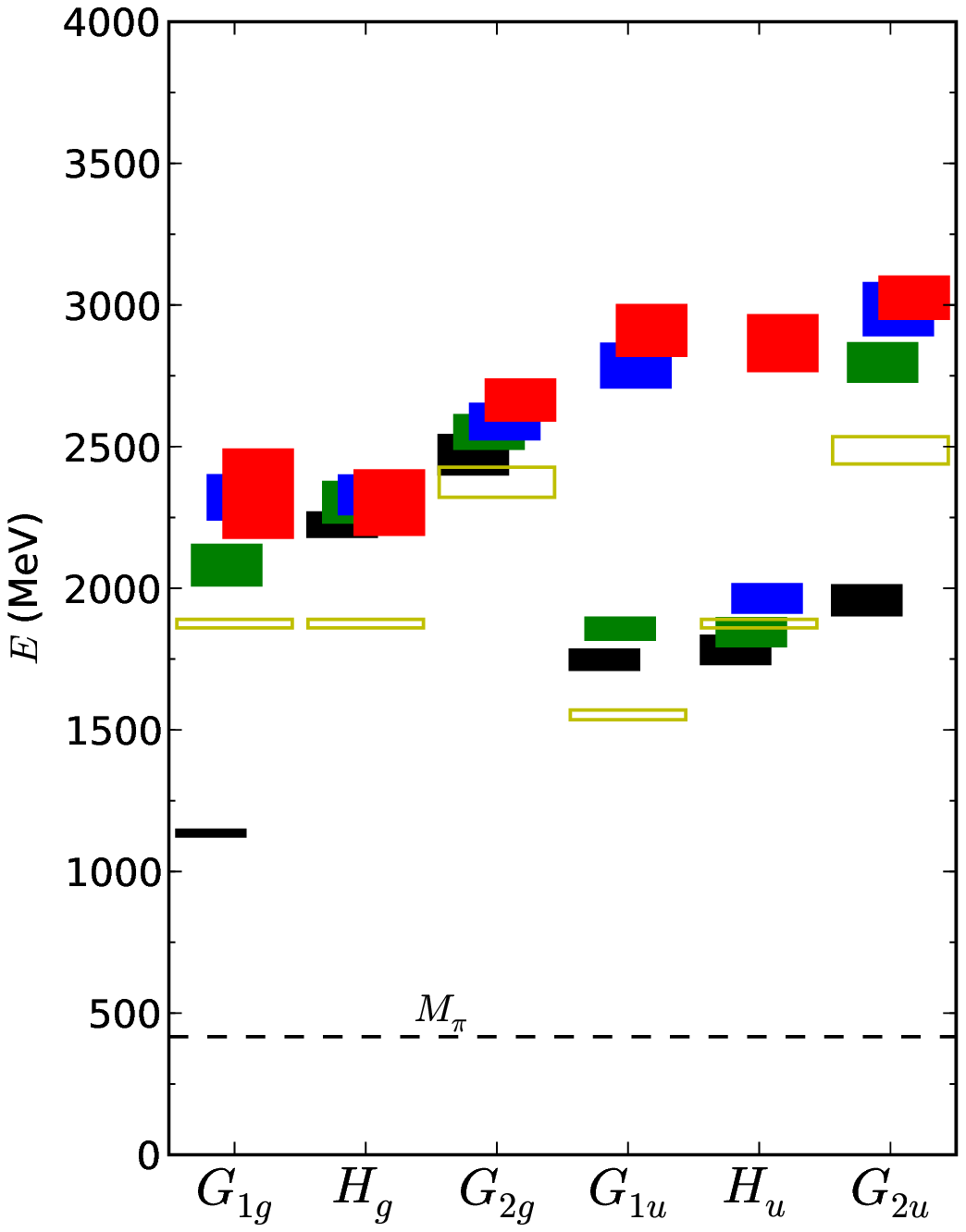} &
\includegraphics[width=0.4\textwidth,clip=true,height=0.5\textwidth]{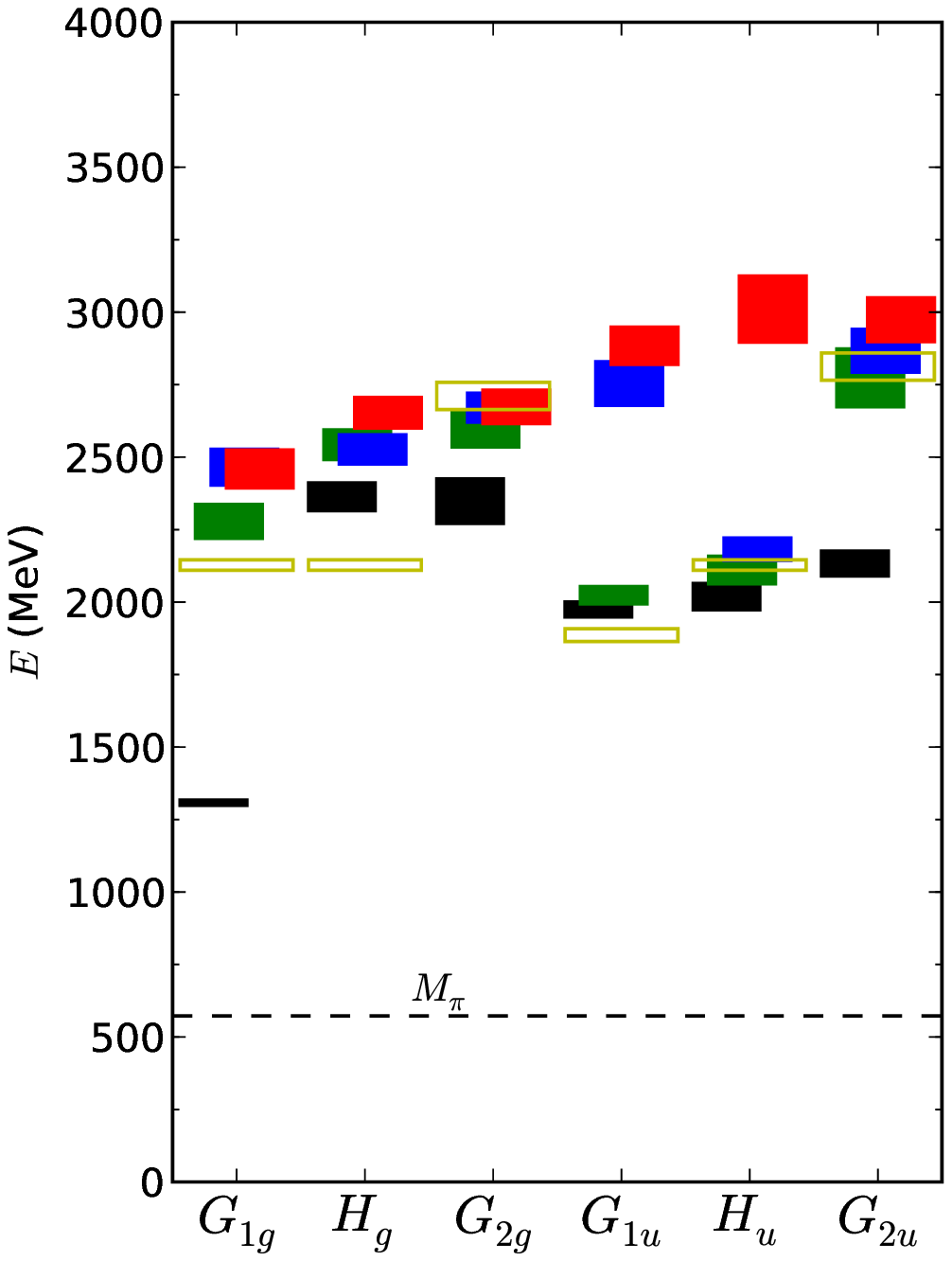}
\end{tabular}
\end{center}
\caption{The energies obtained for each symmetry channel of
isospin $\frac{1}{2}$ baryons are shown based on the
$24^3\times64$ $N_f=2$ lattice QCD data for $m_{\pi}$ = 416 
MeV (left panel) and $m_{\pi}$ = 578 MeV (right panel). 
The scale shows energies in MeV and errors are indicated by the vertical
size of the boxes. The overall error in the scale setting is not included. Empty boxes show thresholds for multi-hadron states.} \label{fig:boxplot_m416_m578}
\end{figure}

\begin{figure}
\begin{tabular}{cc}
\includegraphics[width=0.4\textwidth,height=2in,clip=true]{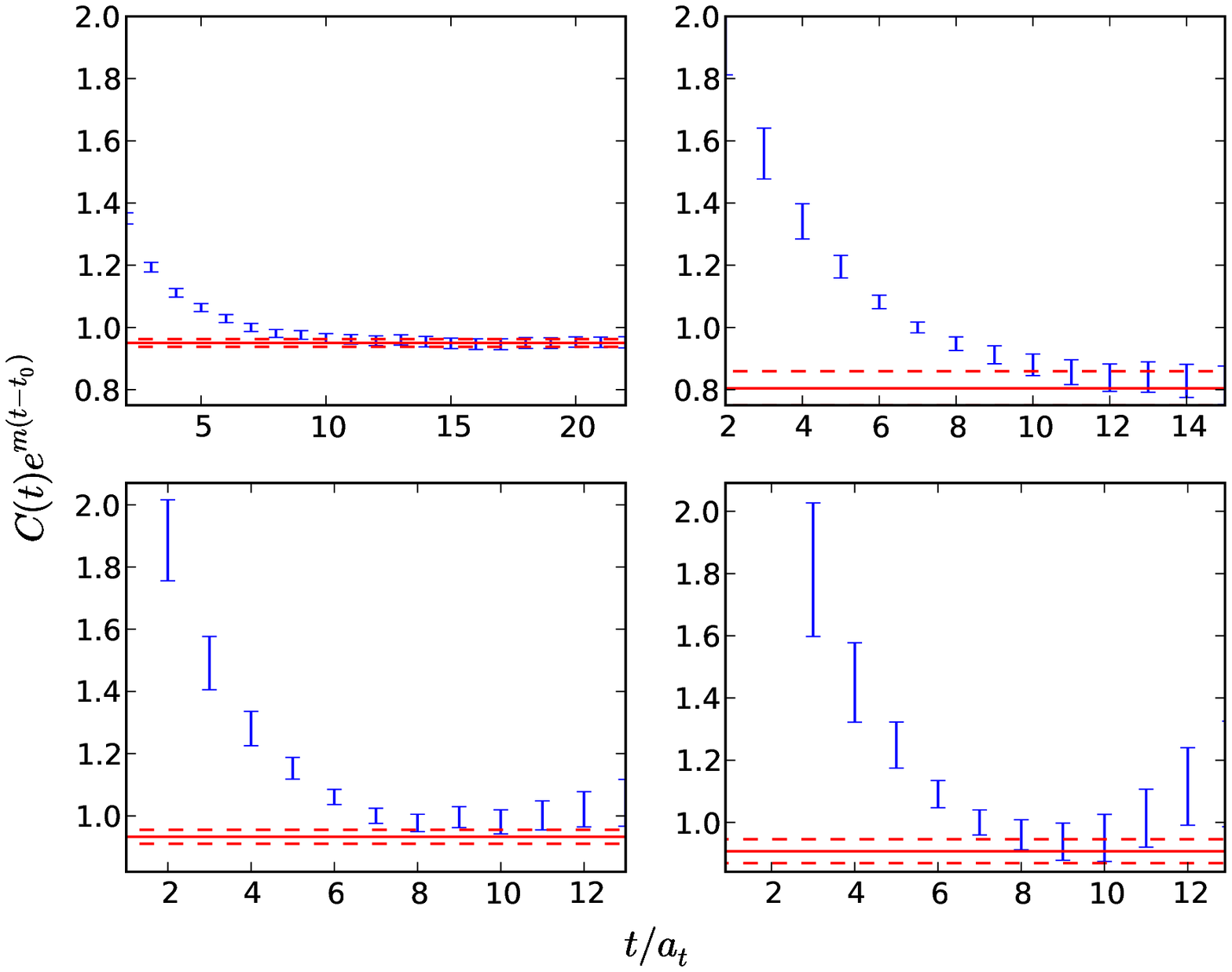}
\includegraphics[width=0.4\textwidth,height=2in,clip=true]{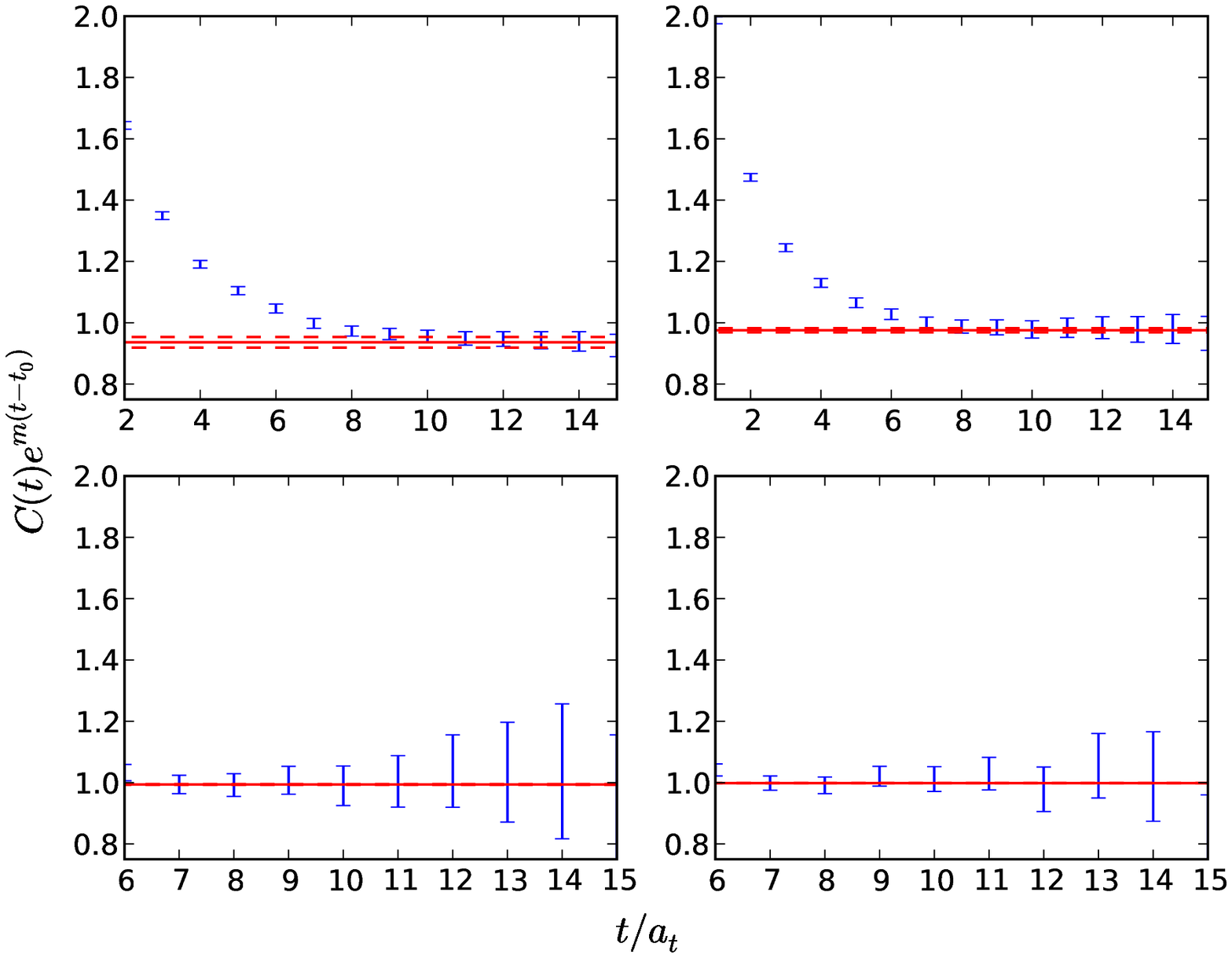}
\end{tabular}
\caption{Plots of Eq.~(\ref{eq:corr}) versus time for $G_{1g}$ states (left panel) and $G_{1u}$ states (right panel)
for $m_{\pi}$=416 MeV. For the two lowest energy states in the $G_{1u}$ channel we first subtract off the backward exponential.}
\label{fig:corr_416_G1}
\end{figure}

\begin{figure}
\begin{tabular}{cc}
\includegraphics[width=0.4\textwidth,height=2in,clip=true]{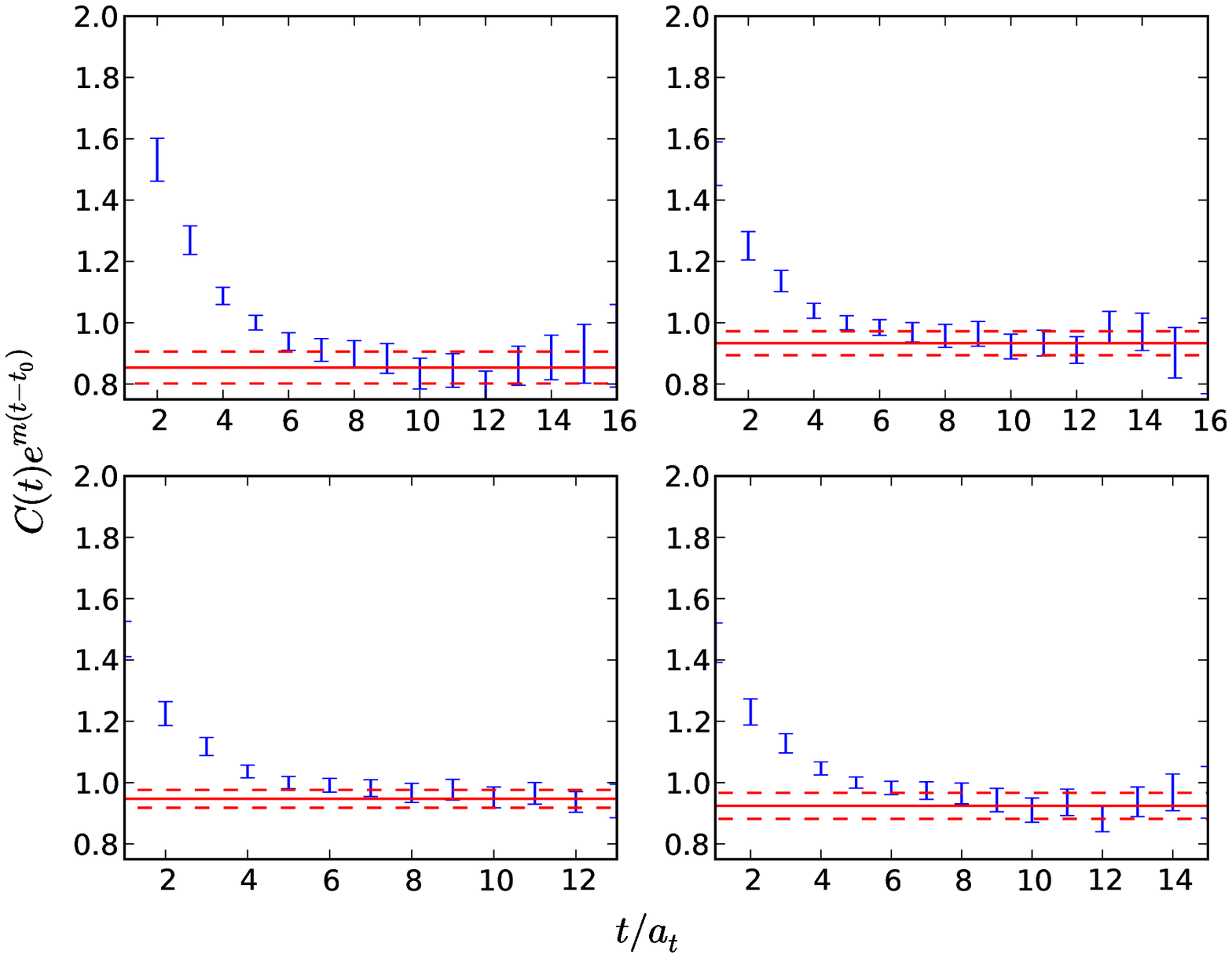}
\includegraphics[width=0.4\textwidth,height=2in,clip=true]{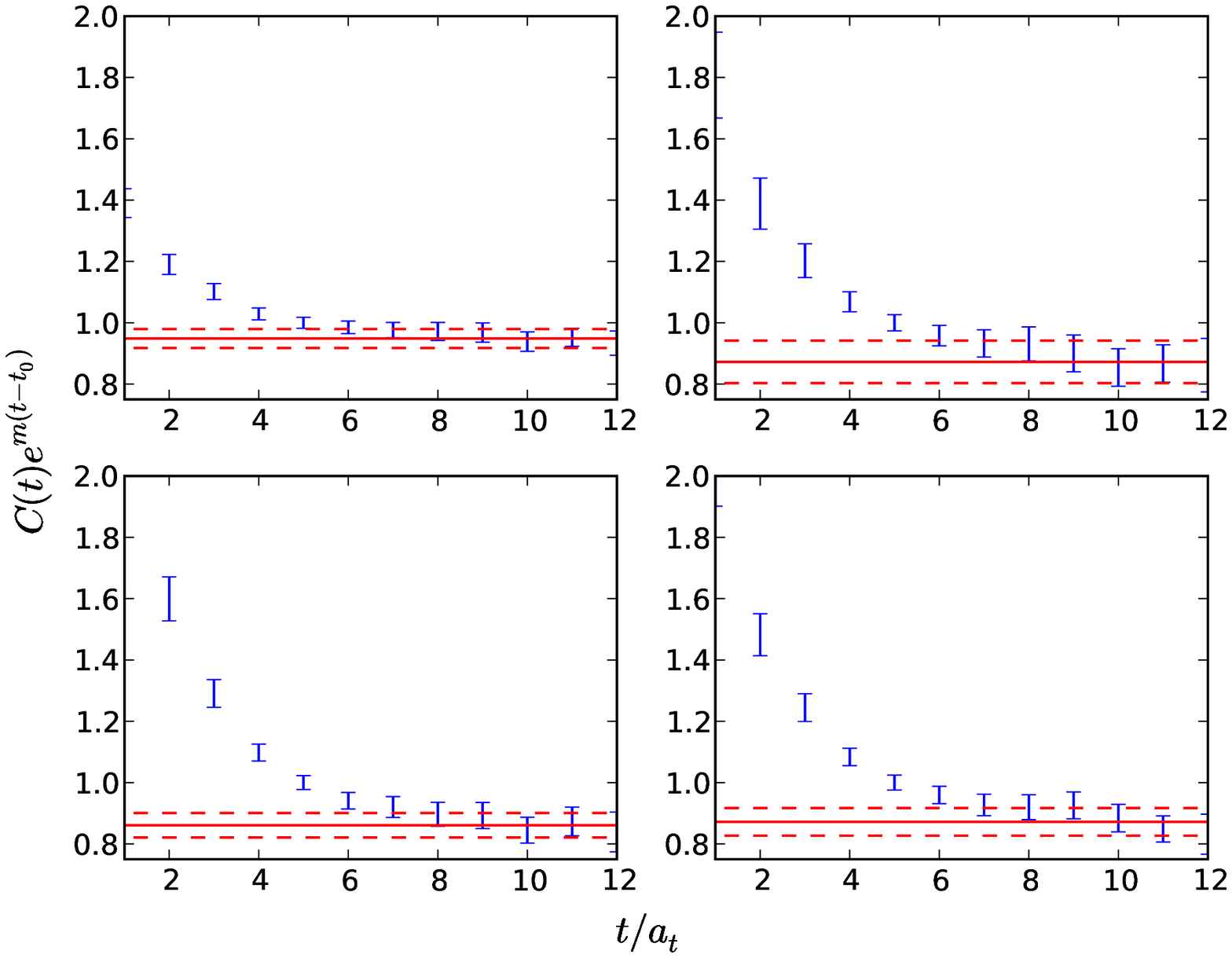}
\end{tabular}
\caption{Plots of Eq.~(\ref{eq:corr}) versus time for $G_{2g}$ states (left panel) and $G_{2u}$ states (right panel)
for $m_{\pi}$=578 MeV.}
\label{fig:corr_578_G2}
\end{figure}

We extracted spectra using the fixed-eigenvector method from 
the $m_{\pi}$ = 416 MeV lattice using 430 gauge configurations
and from the $m_{\pi}$ = 578 MeV lattice using 363 gauge configurations.  
Four states are reported for each channel for both pion masses.
The results 
for $m_{\pi}$ = 416 MeV are given 
in Table \ref{tab:results_416}
and the results for $m_{\pi}$ = 578 MeV are given 
in Table \ref{tab:results_578}.  
The results are based on 16$\times$16 matrices of correlation functions
using values of $t_0$, $t^*$ and the fitting windows $t_i-t_f$ as shown in the tables.
Plots of the $N_f =2$ spectrum for the two $m_{\pi}$ values are
shown in Fig. \ref{fig:boxplot_m416_m578}. 
The pion mass 
is shown by the dashed line and thresholds for multiparticle states
(to be discussed further on) are shown by empty boxes.  Plots of 
Eq.~(\ref{eq:corr}) versus time for each extracted state are shown in 
Figs.~\ref{fig:corr_416_G1}-\ref{fig:corr_578_G2}.


In the positive parity channels, we identify the $G_{1g}$
ground state as the nucleon.  
The spectrum for 
$m_{\pi}$ = 416 MeV is shifted toward higher 
energy values for $m_{\pi}$ = 578 MeV.  
The nucleon mass increases 172 MeV from 1136 MeV to 1308 MeV
when $m_{\pi}$ increases 162 MeV.  
If we extrapolate the nucleon to the physical pion mass using 
the formula $M = a + b m_{\pi}^2$, the result is 972(28) MeV.
 
     Results for the negative-parity excited states 
exhibit some interesting features.
The pattern of $G_{1u}$ energies shows two states at approximately 1.5
and 1.6 times the nucleon mass with the next state much higher.  This 
  pattern is similar to the pattern of masses
of the physical spectrum, which has 
$\frac{1}{2}^-$ resonances at 1535 MeV and 1650 MeV 
with the third $\frac{1}{2}^-$ resonance
well above them at 2090 MeV.  
Because our baryon operators do not 
contain multi-hadron operators, they are expected to couple more strongly to
three-quark states, suggesting that the lowest $G_{1u}$ state is more likely to be
a $N^*$ state.  However, it is above the threshold for a $\pi N$ scattering state so 
further analysis clearly is needed to confirm this assignment.

An isolated state in the $H_u$ irrep corresponds to a spin $\frac{3}{2}^-$ state   
for which the lowest physical state is the $N(1520)$ resonance and the
next to lowest is the $N(1700)$.
In the $H_{u}$ channel, the energies of the three lowest states
are about 1.57, 1.62 and 1.73 times the nucleon mass at the lower pion mass.
 The physical states
for spin $\frac{3}{2}^-$ are 1.62 and 1.81 times the physical nucleon mass.
In the $G_{2u}$ channel at $m_{\pi}$ = 416 MeV, we see that the
lowest-energy state at 1957(51) MeV is degenerate (within errors) 
with the third $H_{u}$ state at 1964(48) MeV with no state at the
same energy in the $G_{1u}$ channel.
A similar pattern 
is seen for $m_{\pi}$ = 578 MeV, except shifted upward by 
about 190 MeV.  The lowest $G_{2u}$ state at 2133(43) MeV is
degenerate with the third $H_u$ state at 2182(38) MeV.  This pattern is the
signature of a spin $\frac{3}{2}^-$ state and a nearby spin $\frac{5}{2}^-$ state.
One $H_u$ state, most likely the second, is the spin $\frac{3}{2}^-$ state and the 
other $H_u$ state is the partner state of the $G_{2u}$ state required for a 
spin $\frac{5}{2}^-$ state.  
The lowest possible spin in $G_{2u}$ is $\frac{5}{2}$ and because the 
$G_{2u}$ irrep has only two of the 2J+1=6 components needed for spin-$\frac{5}{2}$, 
the other four components necessarily are in a partner
$H_u$ state.  For a spin-$\frac{5}{2}$ state, the $G_{2u}$ and
$H_u$ states must be degenerate in the continuum limit and for
a clean interpretation there should not be a $G_{1u}$ state that is degenerate with these
two because that would be the signature  of an isolated
spin-$\frac{7}{2}$ state or a possible accidental degeneracy of a 
spin $\frac{1}{2}$ and $\frac{5}{2}$ states.   Our spectra show evidence for
a spin-$\frac{5}{2}^-$ state and a
spin $\frac{3}{2}$ state close to the same energy.  As the pion mass is reduced to 
140 MeV and the lattice spacing is extrapolated to zero, the partner $H_u$ and $G_{2u}$ states in the lattice spectrum
should approach the lowest $\frac{5}{2}^-$ state in the physical 
spectrum, i.e., $N(1675)$ with a half-width of 75 MeV.  The first and second $H_u$
states should approach the 1520 MeV and 1700 MeV spin $\frac{3}{2}^-$ states in 
the physical spectrum.  

The first excited positive-parity state in $G_{1g}$ is at 2082(70) MeV
for the lighter pion mass.  That is 1.83 times the mass of the  
lowest $G_{1g}$ state (nucleon) and about 334 MeV more massive than the 
lowest $G_{1u}$ state. 
It also is well above the threshold energy for a p-wave $N\pi$ state(1785 MeV 
at the 416 MeV pion mass and this lattice length).
In the physical spectrum the 
first excited, even-parity resonance is $N(1440)\frac{1}{2}^+$ with 
energy 1.53 times the nucleon mass and
below that of the lowest odd-parity $N^*(1535) \frac{1}{2}^-$ state. 
Whether the energy of the first excited $G_{1g}$ state will decrease 
toward the Roper state at lower values of the pion mass remains an open question.

\begin{table}
\begin{tabular}{c c}
\hline\hline
$G_{1g}$, $t_0=7$, $t^*=10$ & $G_{1u}$, $t_0=7$, $t^*=9$ \\
\begin{tabular}{c c c}
 \hline
time & $Ea_t$ & $E$ (MeV)\\
$ 3-21$ & $0.2044(18)$ & $1136(10)$\\
$ 2-14$ & $0.3747(126)$ & $2082(70)$\\
$ 2-12$ & $0.4177(137)$ & $2321(76)$\\
$ 2-12$ & $0.4201(277)$ & $2334(154)$\\
\end{tabular}
&
\begin{tabular}{c c c}
\hline
time & $Ea_t$ & $E$ (MeV) \\
$ 3-14$ & $0.3146(61)$ & $1748(34)$\\
$ 2-14$ & $0.3343(67)$ & $1857(37)$\\
$ 7-14$ & $0.5014(136)$ & $2786(76)$\\
$ 7-13$ & $0.5238(158)$ & $2910(88)$\\
\end{tabular}\\
$H_{g}$, $t_0=8$, $t^*=10$ & $H_{u}$, $t_0=8$, $t^*=9$ \\
\begin{tabular}{c c c}
 \hline
time & $Ea_t$ & $E$ (MeV) \\
$ 3-16$ & $0.4004(74)$ & $2225(41)$\\
$ 3-17$ & $0.4146(126)$ & $2304(70)$\\
$ 3-18$ & $0.4193(120)$ & $2330(67)$\\
$ 3-16$ & $0.4144(202)$ & $2302(112)$\\
\end{tabular}
&
\begin{tabular}{c c c}
\hline
time & $Ea_t$ & $E$ (MeV) \\
$ 3-23$ & $0.3208(87)$ & $1782(48)$\\
$ 3-21$ & $0.3320(86)$ & $1845(48)$\\
$ 3-19$ & $0.3535(87)$ & $1964(48)$\\
$ 2-11$ & $0.5157(174)$ & $2865(97)$\\
\end{tabular}\\
$G_{2g}$, $t_0=6$, $t^*=8$ & $G_{2u}$, $t_0=6$, $t^*=9$ \\
\begin{tabular}{c c c}
 \hline
time & $Ea_t$ & $E$ (MeV) \\
$ 2-12$ & $0.4448(122)$ & $2471(68)$\\
$ 2-12$ & $0.4593(104)$ & $2552(58)$\\
$ 2-11$ & $0.4659(110)$ & $2589(61)$\\
$ 2-14$ & $0.4796(127)$ & $2665(71)$\\
\end{tabular}
&
\begin{tabular}{c c c}
\hline
time & $Ea_t$ & $E$ (MeV) \\
$ 2-17$ & $0.3523(92)$ & $1957(51)$\\
$ 2-12$ & $0.5035(119)$ & $2797(66)$\\
$ 2-12$ & $0.5373(162)$ & $2985(90)$\\
$ 2-10$ & $0.5446(131)$ & $3026(73)$\\
\end{tabular}\\
\hline\hline
\end{tabular}
\caption{Isospin $\frac{1}{2}$ spectrum for $m_{\pi}$ =416 MeV.
The energies in MeV units are based on the scale 
$a_t^{-1} =$ 5556 MeV, and do not include the error in the
the determination of the scale that acts as an overall
multiplicative factor in the range 0.94 to 1.06. 
}
\label{tab:results_416} 
\end{table}

\begin{table}
\begin{tabular}{c c}
\hline\hline
$G_{1g}$, $t_0=6$, $t^*=10$ & $G_{1u}$, $t_0=6$, $t^*=9$ \\
\begin{tabular}{c c c}
 \hline
time & $Ea_t$ & $E$ (MeV)\\
$ 2-27$ & $0.2463(17)$ & $1308(9)$\\
$ 2-15$ & $0.4291(110)$ & $2279(58)$\\
$ 2-15$ & $0.4643(116)$ & $2465(62)$\\
$ 2-11$ & $0.4631(123)$ & $2459(65)$\\
\end{tabular}
&
\begin{tabular}{c c c}
\hline
time & $Ea_t$ & $E$ (MeV) \\
$ 2-11$ & $0.3719(48)$ & $1975(25)$\\
$ 2-11$ & $0.3811(56)$ & $2024(30)$\\
$ 2-11$ & $0.5186(141)$ & $2754(75)$\\
$ 2-11$ & $0.5431(121)$ & $2884(64)$\\
\end{tabular}\\
$H_{g}$, $t_0=6$, $t^*=9$ & $H_{u}$, $t_0=5$, $t^*=7$ \\
\begin{tabular}{c c c}
 \hline
time & $Ea_t$ & $E$ (MeV) \\
$ 2-14$ & $0.4450(90)$ & $2363(48)$\\
$ 2-11$ & $0.4789(96)$ & $2543(51)$\\
$ 2-11$ & $0.4758(95)$ & $2526(50)$\\
$ 2-11$ & $0.4996(99)$ & $2653(53)$\\
\end{tabular}
&
\begin{tabular}{c c c}
\hline
time & $Ea_t$ & $E$ (MeV) \\
$ 2-11$ & $0.3802(86)$ & $2019(46)$\\
$ 2-11$ & $0.3975(89)$ & $2111(47)$\\
$ 2-11$ & $0.4110(72)$ & $2182(38)$\\
$ 2-11$ & $0.5670(215)$ & $3011(114)$\\
\end{tabular}\\
$G_{2g}$, $t_0=5$, $t^*=9$ & $G_{2u}$, $t_0=5$, $t^*=9$ \\
\begin{tabular}{c c c}
 \hline
time & $Ea_t$ & $E$ (MeV) \\
$ 2-15$ & $0.4422(144)$ & $2348(76)$\\
$ 2-15$ & $0.4887(113)$ & $2595(60)$\\
$ 2-12$ & $0.5030(94)$ & $2671(50)$\\
$ 2-14$ & $0.5035(108)$ & $2674(57)$\\
\end{tabular}
&
\begin{tabular}{c c c}
\hline
time & $Ea_t$ & $E$ (MeV) \\
$ 2-11$ & $0.4017(81)$ & $2133(43)$\\
$ 2-11$ & $0.5223(188)$ & $2773(100)$\\
$ 2-11$ & $0.5399(139)$ & $2867(74)$\\
$ 2-11$ & $0.5601(142)$ & $2974(75)$\\
\end{tabular}\\
\hline\hline
\end{tabular}
\caption{Isospin $\frac{1}{2}$ spectrum for $m_{\pi}$ =578 MeV.
The energies in MeV units are based on the scale 
$a_t^{-1} = $ 5310 MeV, and do not include the error in the
scale determination that acts as an overall multiplicative factor
in the range 0.95 to 1.05. 
}
\label{tab:results_578} 
\end{table}

A signal for a $\frac{5}{2}^-$ state could not be clearly identified
in the quenched QCD analysis of Ref.~\cite{Basak:2007kj} at 480 MeV pion mass.  That spectrum
had larger errors and showed three degenerate states 
(within errors) in the $G_{2u}$, $H_u$ and $G_{1u}$ irreps, a pattern with two 
possible interpretations.  It could be a
single spin-$\frac{7}{2}^-$ state
or an accidental degeneracy of a spin-$\frac{5}{2}^-$ state and a spin-$\frac{1}{2}^-$ state.
For $N_f=2$ QCD and $m_{\pi}$ = 416 and 578 MeV, we see clear evidence 
for a $\frac{5}{2}^-$ state.

\begin{figure}
\begin{tabular}{cc}/

\includegraphics[width=0.4\textwidth,height=2in,clip=true]{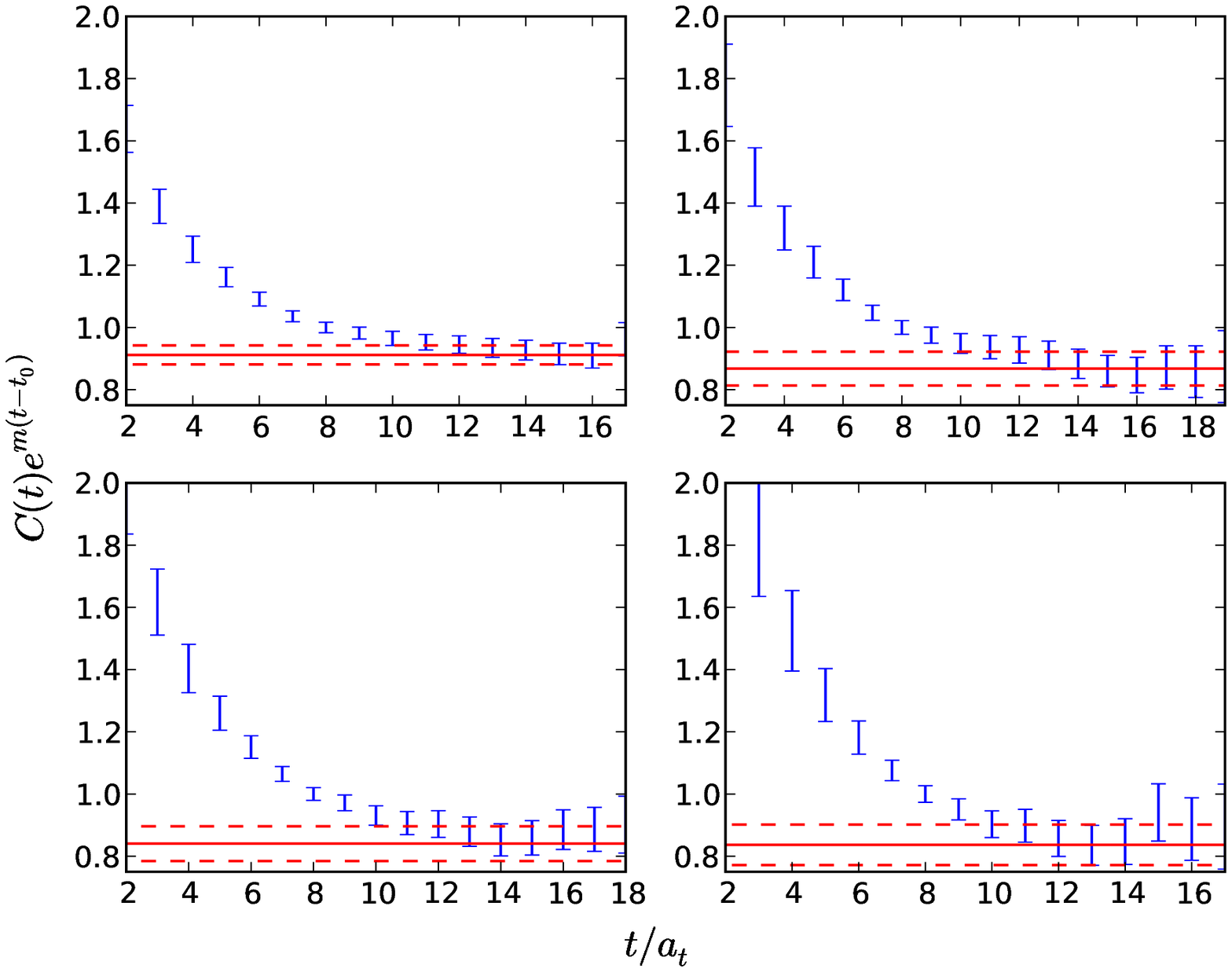}
\includegraphics[width=0.4\textwidth,height=2in,clip=true]{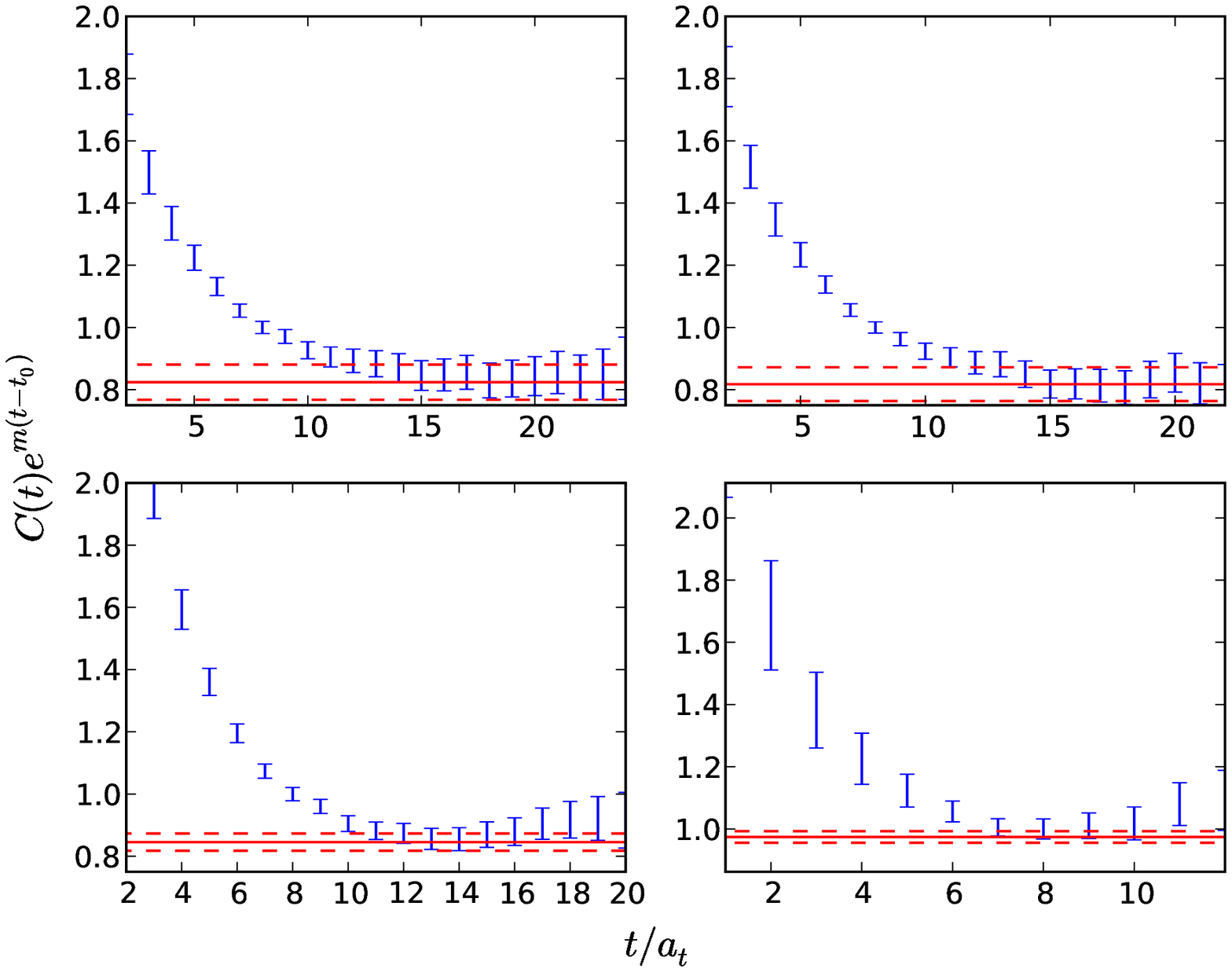}
\end{tabular}
\caption{Plots of Eq.~(\ref{eq:corr}) versus time for $H_g$ states (left panel) and $H_u$ states (right panel)
for $m_{\pi}$=416 MeV.}
\end{figure}

\begin{figure}
\begin{tabular}{cc}
\includegraphics[width=0.4\textwidth,height=2in,clip=true]{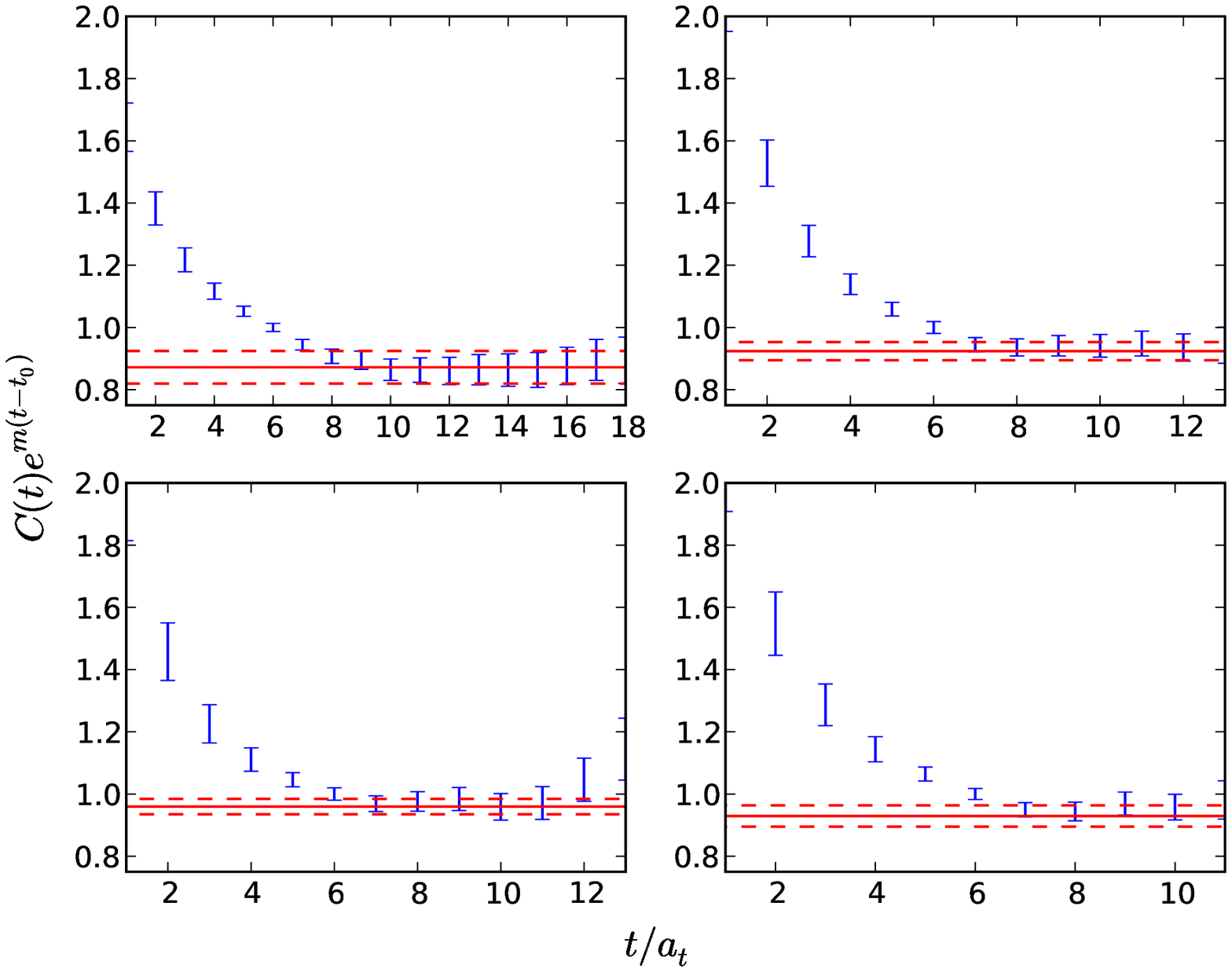}
\includegraphics[width=0.4\textwidth,height=2in,clip=true]{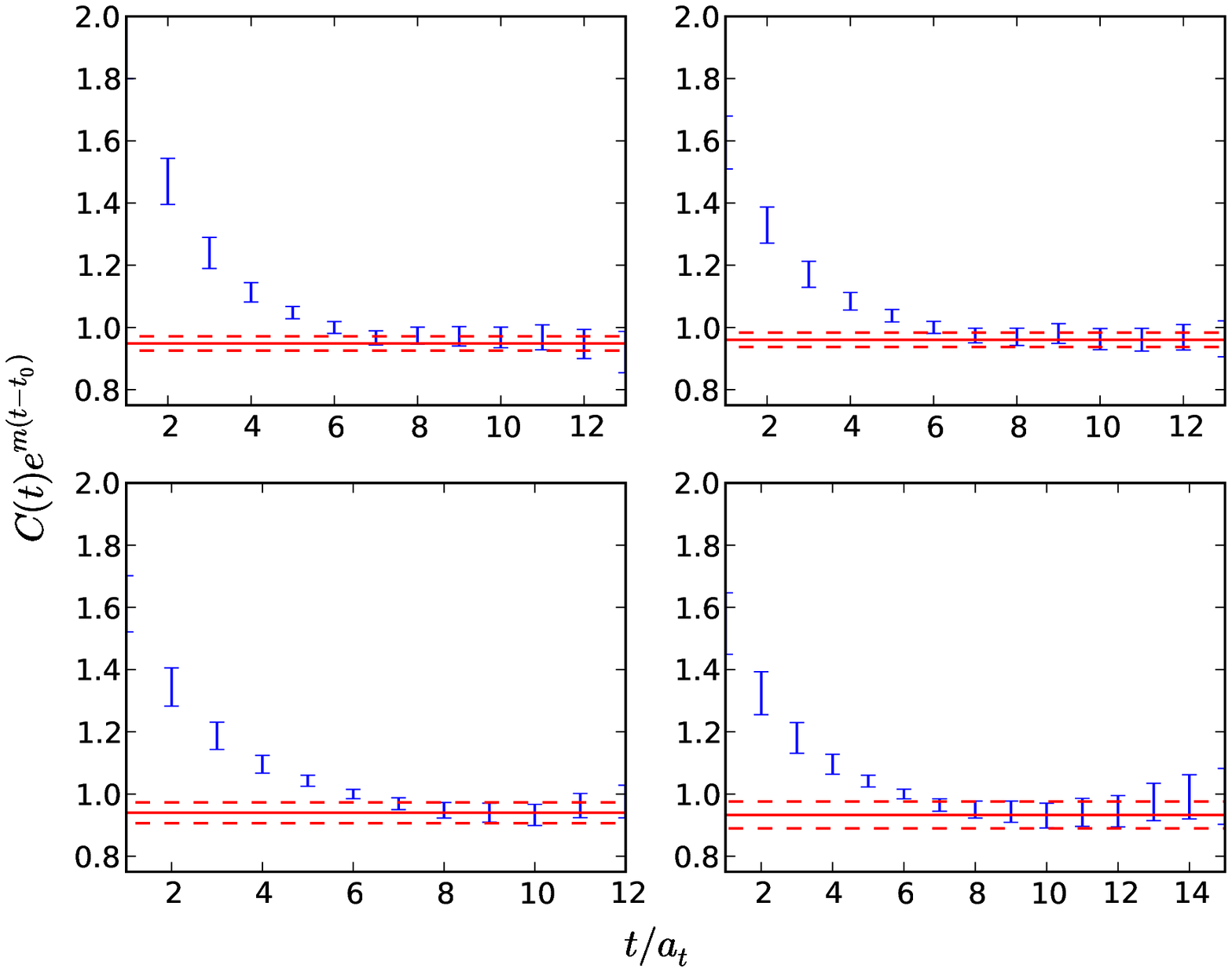}
\end{tabular}
\caption{Plots of Eq.~(\ref{eq:corr}) versus time for $G_{2g}$ states (left panel) and $G_{2u}$ states (right panel)
for $m_{\pi}$=416 MeV.}
\end{figure}

\begin{figure}
\begin{tabular}{cc}
\includegraphics[width=0.4\textwidth,height=2in,clip=true]{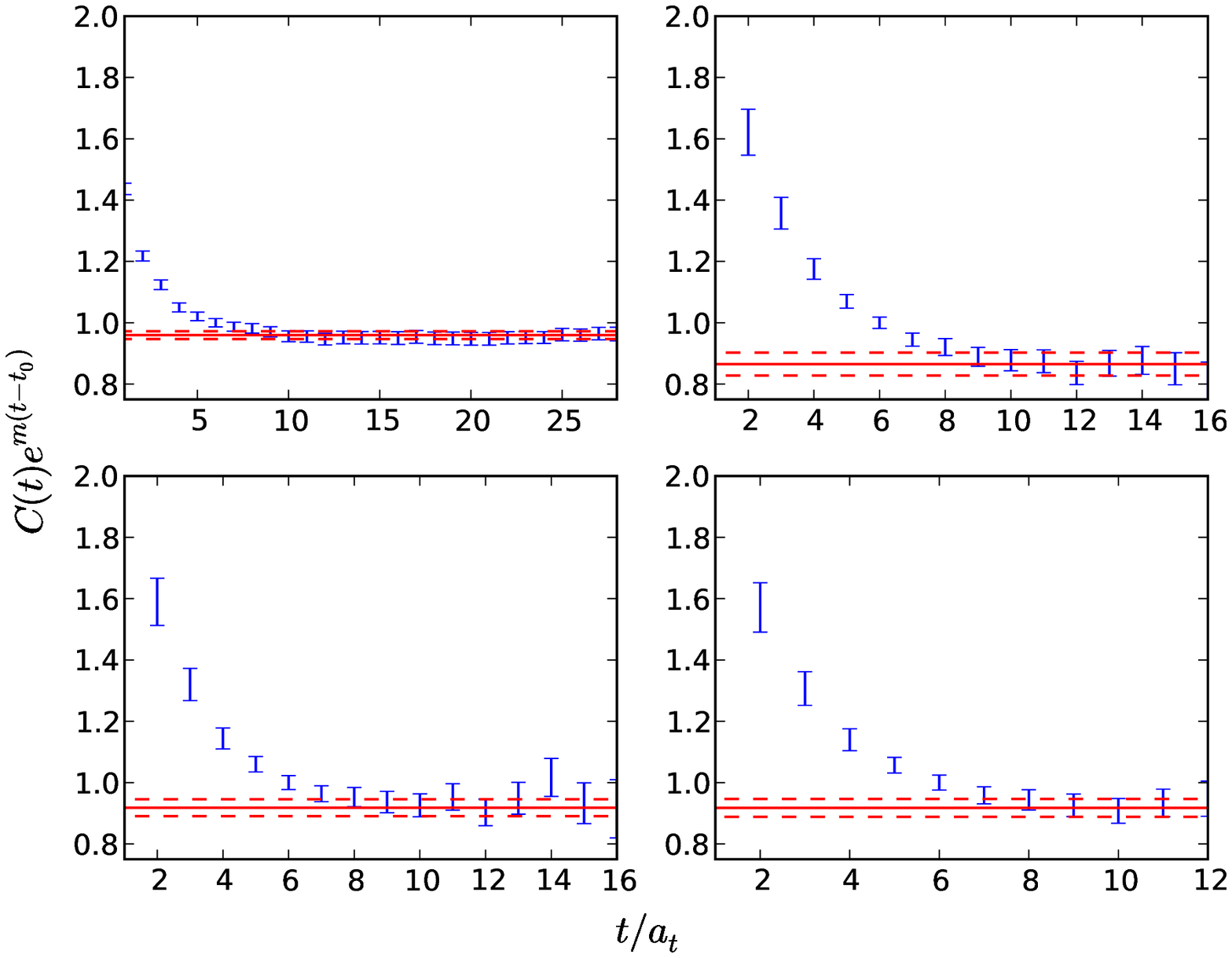}
\includegraphics[width=0.4\textwidth,height=2in,clip=true]{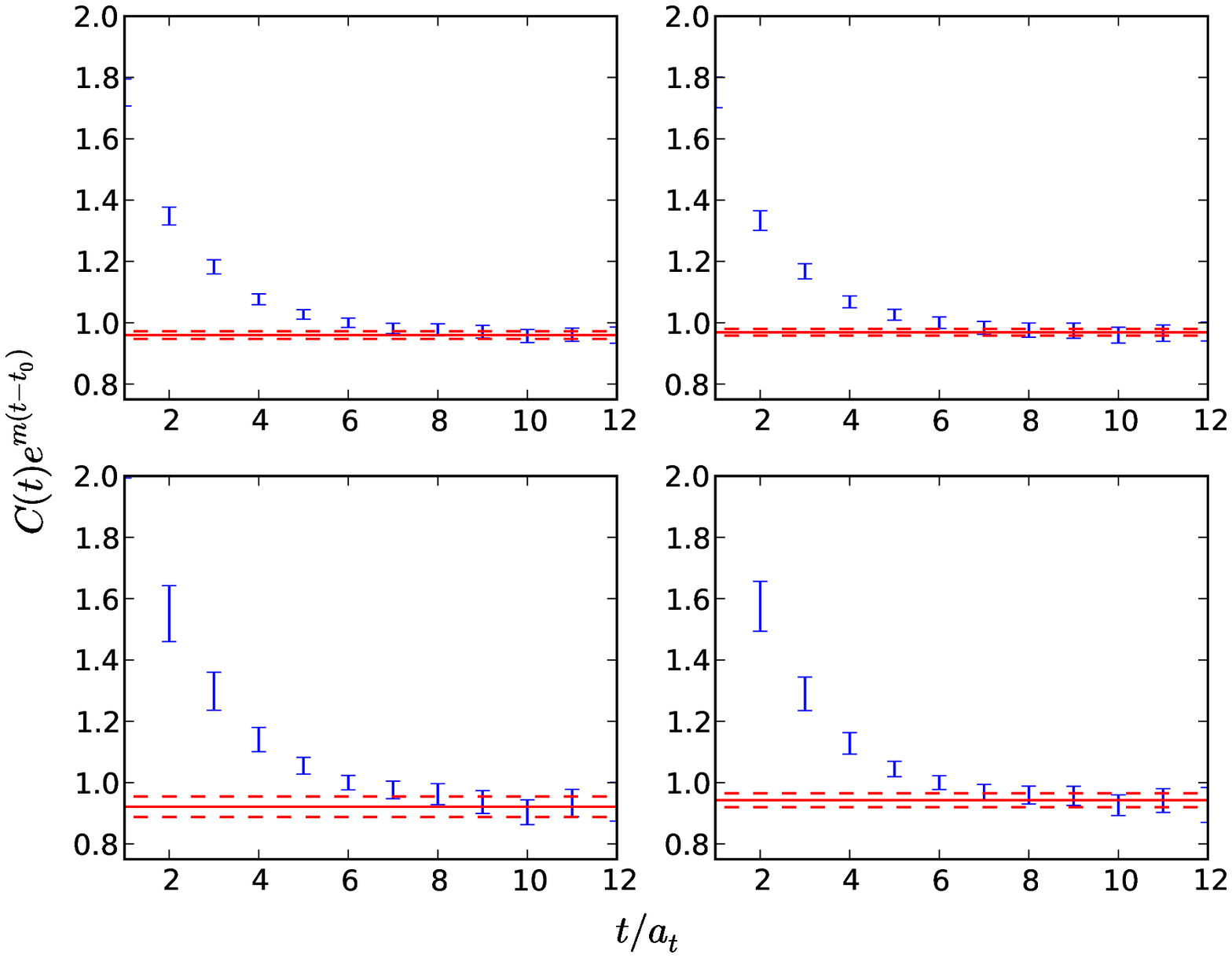}
\end{tabular}
\caption{Plots of Eq.~(\ref{eq:corr}) versus time for $G_{1g}$ states (left panel) and $G_{1u}$ states (right panel)
for $m_{\pi}$=578 MeV.}
\end{figure}

\begin{figure}
\begin{tabular}{cc}
\includegraphics[width=0.4\textwidth,height=2in,clip=true]{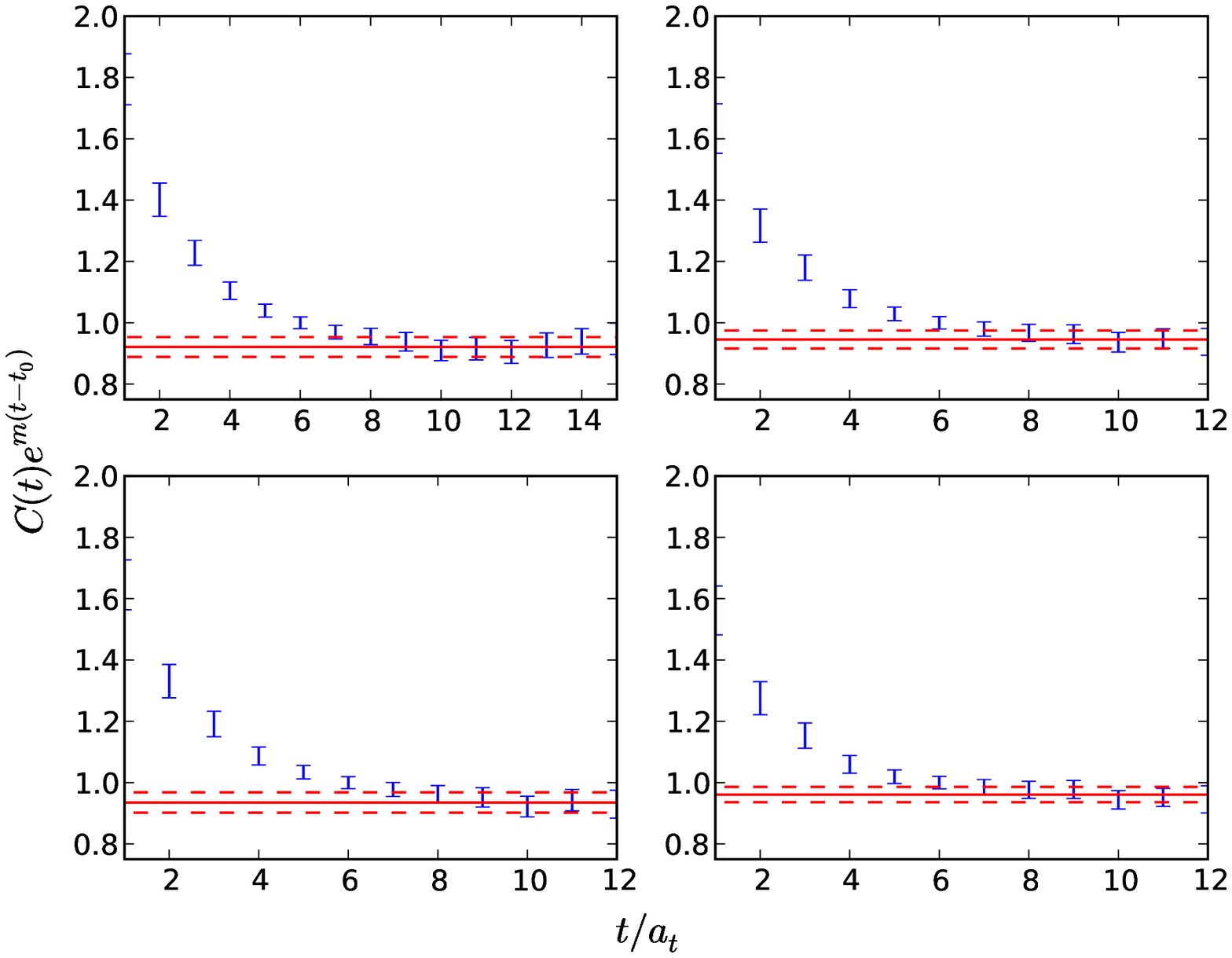}
\includegraphics[width=0.4\textwidth,height=2in,clip=true]{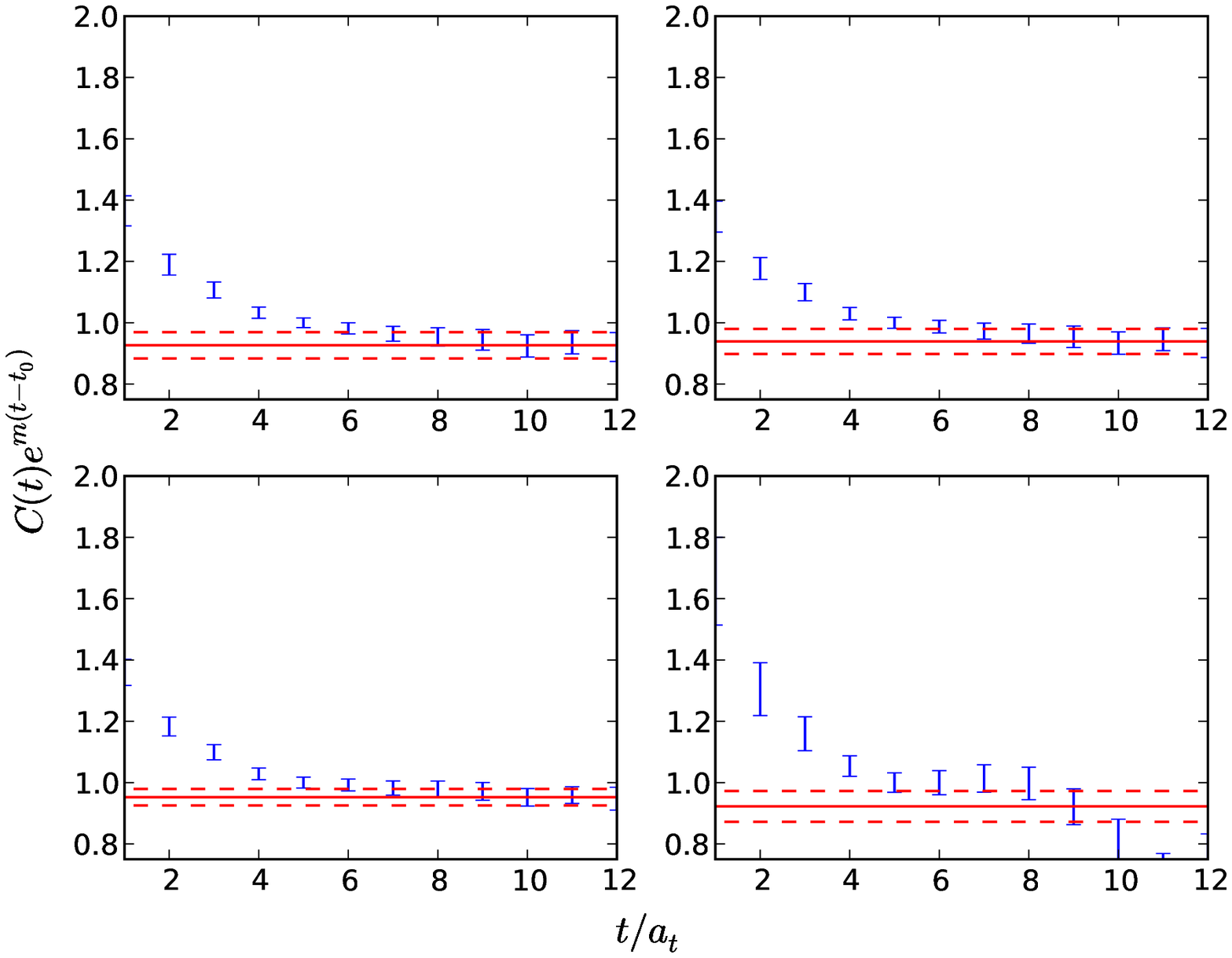}
\end{tabular}
\caption{Plots of Eq.~(\ref{eq:corr}) versus time for $H_g$ states (left panel) and $H_u$ states (right panel)
for $m_{\pi}$=578 MeV.}
\end{figure}

%% file: multihadrons.tex

As the pion mass decreases,
it becomes increasingly likely that 
some of the energy levels determined in our simulations 
will correspond to multi-hadron states. 
Disentangling these states from the 
hadron spectrum
will be challenging and 
will require the use of specially-designed multi-hadron operators. 
In this paper, as a first step towards the identification of scattering states, we estimate 
multi-hadron threshold energies in each of the irreducible representations of $O_{h}$.
Some of the threshold energies correspond to states with two hadrons at rest. 
However, scattering states of hadrons with back to back momenta must also be considered.

On the lattice, a hadron with momentum $\vec{p}$ 
transforms irreducibly under the space group, which is the
semi-direct product of the group of three-dimensional lattice 
translations with $O_{h}$.
In addition to the momentum vector $\vec{p}$, irreducible representations 
of the lattice space group are characterized by 
a label denoting the irreducible representations of the group of 
lattice rotations which leaves $\vec{p}$ invariant (the little group of 
$\vec{p} $). 
For particles at rest, the little group is $O_{h}$.
More generally, the little group is a subgroup of $O_{h}$ which
depends on the orientation of $\vec{p}$ with respect to the lattice axes.
The minimum non-zero momenta 
on a periodic lattice, of magnitude $ 2 \pi/ \left(N_{s} a_{s} \right) $, 
are directed along 
the lattice axes.
The little group for such momenta is $C_{4 \nu}$.

Given the spectrum of hadrons at rest, 
one can deduce the allowed free-particle 
energies in any irreducible representation of the space group. 
To see this, we first note 
that representations of a lattice 
little group can be subduced from the irreducible representations of 
$O_{h}$. The subduced representations are in general reducible and may be decomposed 
into a direct sum of irreducible little group representations. 
Irreducible representations of the full space group are induced 
from the irreducible representations of the lattice little groups. 
Thus, one can relate the  irreducible representations of the space group 
to the representations of $O_{h}$.
Neglecting cutoff effects, the energy of a non-interacting hadron 
with momentum $\vec{p}$ is given by $ E = \sqrt{M^{2}_{h} + |\vec{p}|^{2}}$, 
where $M_{h}$ is the rest mass of the hadron. Therefore, provided that the 
hadron rest masses are known, the free-particle energies in representations 
with non-zero $\vec{p}$ can be determined.

Ref.~\cite{Moore:2006ng} gives the decomposition of direct products of irreducible 
representations of the space group, including representations 
with non-zero momentum, into the irreducible representations of $O_{h}$. 
We use this information to identify the allowed multi-hadron states in 
each representation of $O_{h}$. The energies of multi-hadron states 
are approximated by the sum of the energies of their 
constituents. The empty boxes in Fig. \ref{fig:boxplot_m416_m578}  
show
candidates for multi-hadron thresholds for both pion masses.
Note that $I= \frac{3} {2}$ baryons, which are not considered 
in this study, can also combine with isovector mesons to form $I = \frac{1} {2}$ two-particle states. 
However, such states are expected to lie above the thresholds presented here.
In both figures, the threshold energies in the $G_{1 u}$ and $G_{2g}$ representations correspond to meson-baryon states 
involving a pion at rest, while the other thresholds involve particles with non-zero momentum.
The threshold energies in the $G_{1 g}$, $H_{g}$ and $H_{u}$ representations are degenerate.
Our results illustrate the need for a proper analysis of multi-hadron contamination.
Even at the heavier pion mass, many of the measured energy levels lie above the threshold for scattering states.
Due to lattice artifacts, finite volume effects and the interaction between hadrons, the measured multi-hadron 
energies are expected to deviate from our estimates. 
This might explain some of the discrepancies between the predicted multi-hadron energies and the measured spectrum. 
However, it is also likely that the interpolating operators used in our simulations, selected on the basis of a quenched study, 
couple only weakly to the lowest-lying multi-hadron states.
Nevertheless, our analysis indicates that 
multi-hadron states cannot be discounted, even at the moderate pion masses used in this study.


%% file: conclusions.tex
\section{Conclusion and Outlook}\label{Sec:Conclusion}
 In this work, anisotropic lattices with $a_t = \frac{1}{3}a_s$ are developed
for $N_f =2$ QCD with two pion masses: $m_{\pi}=$ 416 MeV and 578 MeV.
The lattice setup and the algorithms used to generate gauge configurations
are described in detail.  Conventional two-point correlation functions
are used to calculate the spectrum of mesons in order to determine the
pion masses and to tune the fermion anisotropy to $\xi = 3$, which matches that
of the gauge fields.  The lattice scales $a_s \approx$ 0.113 fm and 0.108 fm
are set using the Sommer parameter.

This work builds upon several years of work to develop large numbers of baryon
operators, to project them to the relevant irreducible representations of the octahedral
group, to optimize the smearing of both the quark and gluon fields in the operators in order
to be able to extract clean signals for effective masses and to prune the
operators to manageable sets of 16 operators that yield good signals for
baryons.  Using the final operators,
16$\times$16 matrices of correlation functions are calculated
in each irrep and a variational analysis of the isospin $\frac{1}{2}$ spectrum is
carried out.  The lowest four energy levels in each irrep are reported.
The analysis of the negative-parity spectrum shows a cluster of
states near 1.5 to 1.7 times the nucleon mass that includes a
$\frac{5}{2}^-$ state, two $\frac{1}{2}^-$ states at somewhat lower
energies and two $\frac{3}{2}^-$ states.  This pattern is in accord with
the pattern of physical states, although the latter is at a lower overall
energy scale.  The clear signal for a $\frac{5}{2}^-$ state 
has not been realized previously. The analysis of the
positive-parity spectrum for both pion masses shows that excited states
typically have energies about 1.8 or more times the mass of the nucleon state.
The question remains open whether as the pion mass is reduced the first excited $G_{1g}$ state will
come down to about 1.53 times the nucleon mass, 
where it would agree with the Roper resonance.

All the excited states in the lattice spectrum are near or above
the threshold for $\pi N$ scattering states.  In order to deal properly with
that aspect, multi-hadron operators and all-to-all propagators will
be needed. This is an immediate challenge for progress on the 2+1-flavor dynamical lattices~\cite{Edwards:2008ja,Lin:2008pr}
and it will be addressed in the near future.

\section*{Acknowledgements}
This work was done using the Chroma software suite~\cite{Edwards:2004sx} on clusters at Jefferson Laboratory using time awarded under the USQCD Initiative.
This research used resources of the National Center for Computational Sciences at Oak Ridge National Laboratory, which is supported by the Office of Science of the Department of Energy under Contract DE-AC05-00OR22725.
In particular, we made use of the Jaguar Cray XT facility, using time allocated through the US DOE INCITE program.
This research was supported in part by the National Science Foundation (NSF-PHY-0653315 and NSF-PHY-0510020) through the San Diego Supercomputing Center (SDSC) and the Texas Advanced Computing Center (TACC).
Computational support was provided though Teragrid Resources
provided by the San Diego Supercomputing Center (Blue Gene).
JB, JF and CM were supported by grants NSF-PHY-0653315
and NSF-PHY-0510020;
EE and SW were supported by DOE grant DE-FG02-93ER-40762;
NM was supported under grant No. DST-SR/S2/RJN-19/2007;
AL was supported by RIKEN and Brookhaven National Laboratory under Department of Energy contract
DE-AC02-98CH10886.
EE thanks J. Dudek for help regarding the reconstruction of 
the correlator and for his fitting code.
Authored by Jefferson Science Associates, LLC under U.S. DOE Contract No. DE-AC05-06OR23177. The U.S. Government retains a non-exclusive, paid-up, irrevocable, world-wide license to publish or reproduce this manuscript for U.S. Government purposes.